\documentclass[amssymb,reprint,superscriptaddress,showpacs,longbibliography]{revtex4-1}
\usepackage{amsmath}
\usepackage{amsfonts}
\usepackage{amssymb}
\usepackage{epsfig}
\usepackage{graphicx}

\renewcommand{\phi}{\varphi}

\newcommand{\be}{\begin{equation}}
\newcommand{\ee}{\end{equation}}
\newcommand{\bea}{\begin{equnaray}}
\newcommand{\eea}{\end{equnaray}}
\newcommand{\ba}{\begin{align}}
\newcommand{\ea}{\end{align}}

\usepackage{color}

\definecolor{green}{rgb}{0.0, 0.44, 0.0}
\definecolor{red}{rgb}{1.0, 0.13, 0.32}
\definecolor{blue}{rgb}{0.06, 0.2, 0.65}
\definecolor{magenta}{rgb}{1.0, 0.0, 1.00}
\definecolor{purple}{rgb}{0.7, 0.0, 0.7}
\definecolor{cyan}{rgb}{0.0, 1.0, 1.0}

\begin{document}

\title{Role of fluctuations in the yielding transition of two-dimensional glasses}

\author{Misaki Ozawa}

\affiliation{Laboratoire de Physique de l'Ecole Normale Superieure, ENS, Universite PSL, CNRS, Sorbonne Universite, Universite de Paris, F-75005 Paris,
France}

\author{Ludovic Berthier}

\affiliation{Laboratoire Charles Coulomb, UMR 5221 CNRS-Universit\'e de Montpellier, Montpellier, France}

\affiliation{Department of Chemistry, University of Cambridge, Lensfield Road, Cambridge CB 2 1EW, United Kingdom}

\author{Giulio Biroli}

\affiliation{Laboratoire de Physique de l'Ecole Normale Superieure, ENS, Universite
PSL, CNRS, Sorbonne Universite, Universite de Paris, F-75005 Paris,
France}

\author{Gilles Tarjus}

\affiliation{LPTMC, CNRS-UMR 7600, Sorbonne Universit\'e, 4 Pl. Jussieu, 75005 Paris, France} 

\begin{abstract}
We numerically study yielding in two-dimensional glasses which are generated with a very wide range of stabilities by swap Monte-Carlo simulations and then slowly deformed at zero temperature. We provide strong numerical evidence that stable glasses yield via a nonequilibrium discontinuous transition in the thermodynamic limit. A critical point separates this brittle yielding from the ductile one observed in less stable glasses. 
We find that two-dimensional glasses yield similarly to their three-dimensional counterparts but display larger sample-to-sample disorder-induced fluctuations, stronger finite-size effects, and rougher spatial wandering of the observed shear bands. These findings strongly constrain effective theories of yielding.
\end{abstract}

\maketitle

\section{Introduction}
\label{sec:Intro}

Amorphous solids encompass a wide variety of systems ranging from molecular and metallic glasses to granular media, also including foams, pastes, emulsions, and colloidal glasses. Their mechanical response to a slowly applied deformation exhibits features such as localized plastic rearrangements, avalanche-type of motion, the emergence of strain localization and shear bands~\cite{nicolas2018deformation,barrat2011heterogeneities,rodney2011modeling,bonn2017yield,falk2011deformation}. 
The universality of these phenomena suggests that a unified description  may be possible. When slowly deformed at low temperature from an initial quiescent state, amorphous solids yield beyond some finite level of applied strain and reach a steady state characterized by plastic flow. Understanding yielding is a central issue in materials science, where one would like to avoid the unwanted sudden failure of deformed glass samples~\cite{greer2013shear}. It is also a challenging problem in nonequilibrium statistical physics~\cite{nicolas2018deformation}. 

The ways in which amorphous materials yield can be classified in two main categories: the ``brittle'' extreme where the sample catastrophically breaks into pieces and show macroscopic shear bands~\cite{greer2013shear}, as often observed in molecular and metallic glasses,  and the ``ductile'' behavior in which plastic deformation increases progressively~\cite{bonn2017yield}, commonly found in soft-matter glassy systems. 
The key question is whether the observed variety of yielding behaviors should be described by (i) completely distinct approaches, with, e.g., ductile yielding being describable by soft glassy rheology models~\cite{fielding2000aging} while brittle yielding falls into the realm of the theory of fracture~\cite{alava2006statistical,bouchbinder2014dynamics,popovic2018elastoplastic}; (ii) as a unique phenomenon, taken as the ubiquitous limit of stability of a strained solid in the form of a critical spinodal~\cite{rainone2015following,urbani2017shear,jaiswal2016mechanical,parisi2017shear}, 
or, (iii) as we have recently argued on the basis of mean-field elasto-plastic models and simulations of a three-dimensional atomic glass~\cite{ozawa2018random}, within a unique theoretical framework but with the nature of yielding depending on the degree of effective disorder that is controlled by the preparation of the amorphous solid. 
In the latter case, and in analogy with an athermally driven random-field Ising model (RFIM)~\cite{sethna2006random}, yielding evolves from a mere crossover (for poorly annealed samples) to a nonequilibrium discontinuous transition past a spinodal point (for well-annealed, very stable samples); 
the transition between these two regimes is marked by a critical point that takes place for a specific glass preparation~\cite{nandi2016spinodals} (see also a different approach~\cite{da2020rigidity}). Here we focus on a uniform shear deformation but a similar scenario would apply to an athermal quasi-static oscillatory shear protocol~\cite{yeh2019glass,bhaumik2019role}. Strictly speaking, these sharp transitions can only be observed at zero temperature in strain-controlled quasi-static protocols~\cite{murari2019brittle}. However, temperature is likely to play a minor role in realistic situations given the large energy scales at play. In fact, there is experimental evidence that a given material may indeed show brittle or ductile yielding depending on preparation history of the sample~\cite{shen2007plasticity,kumar2013critical,arif2012spontaneous,yang2012size,ketkaew2018mechanical}.

If yielding is a bona fide (albeit nonequilibrium) discontinuous phase transition ending in a critical point akin to that of a RFIM, one should wonder about its universality class and its dependence on space dimension. By default or with the assumption that the phenomenology is qualitatively unchanged when changing dimension, many of the numerical studies of yielding in model amorphous solids have been carried out in two dimensions (2D)~\cite{shi2005strain,maloney2009anisotropic,barbot2019rejuvenation}. Whereas this may be legitimate when focusing on the flowing steady state or on very ductile behavior,  
one should be more cautious about the role of spatial fluctuations on the nature, and even the existence, of the yielding transition itself as one decreases the dimension of space. Fluctuations of the order parameter are expected to change the values of the exponents of the critical point as dimension is decreased below an  upper critical dimension at which the mean-field description becomes qualitatively valid. More importantly, they smear the transition below a lower critical dimension. In the standard RFIM with ferromagnetic short-range interactions and short-range correlated random fields, this lower critical dimension has been proven to be $D=2$ for the equilibrium behavior~\cite{nattermann1998theory} and, although still debated~\cite{spasojevic2011numerical,balog2018criticality,hayden2019unusual}, appears to also be $D=2$ for the out-of-equilibrium situation of the quasi-statically driven RFIM at zero temperature. However, we expect that the relevant RFIM providing an effective theory for yielding is not the standard one. It is indeed known that elastic interactions in an amorphous solids are long-ranged and anisotropic~\cite{eshelby1957determination,picard2004elastic} instead of short-ranged ferromagnetic, as indeed shown by the appearance of strong anisotropic strain localization in the form of shear bands. 

Therefore, a careful study of yielding in 2D model atomic glasses as a function of preparation is both of fundamental interest and relevant to two-dimensional physical materials, such as dry foams~\cite{arif2012spontaneous}, grains~\cite{ponson2016statistical}, or silica glasses~\cite{huang2013imaging}.
This is what we report in this article, where we consider glass samples that are prepared by optimized swap Monte-Carlo simulations~\cite{ninarello2017models,berthier2019zero} in a wide range of stability from poorly annealed glasses to very stable glasses and that are sheared through an athermal quasi-static protocol. We provide strong evidence that strained 2D stable glasses yield through a sharp discontinuous stress drop, which from finite-size scaling analysis survives in the thermodynamic limit, as in 3D. As the stability of the glass decreases, brittleness decreases and below a critical point, which is characterized by a diverging susceptibility, yielding becomes smooth. Compared to the 3D case, we find that the 2D systems are subject to larger sample-to-sample fluctuations, stronger finite-size effects, and rougher spatial wandering of the shear bands.

This paper is organized as follows.
We describe the simulation methods in Sec.~\ref{sec:methods}.
We demonstrate that two-dimensional stable glasses yield via a nonequilibrium discontinuous transition in Sec.~\ref{sec:discontinuous_transition}.
Then we show that the brittle yielding in 2D displays larger sample-to-sample disorder-induced fluctuations with stronger finite-size effects in Sec.~\ref{sec:anomalous}, and rougher spatial wandering of the observed shear bands in Sec.~\ref{sec:rough}.
Section~\ref{sec:critical} presents a critical point separating the brittle and ductile yielding.
Finally, we conclude our results in Sec.~\ref{sec:conclusions}.

\section{Simulation methods}
\label{sec:methods}

\subsection{Model}

The two-dimensional glass-forming model consists of particles with purely repulsive interactions and a continuous size polydispersity~\cite{ninarello2017models,berthier2019zero}. Particle diameters, $d_i$, are randomly drawn from a distribution of the form: $f(d) = Ad^{-3}$, for $d \in [ d_{\rm min}, d_{\rm max} ]$, where $A$ is a normalization constant. The size polydispersity is quantified by $\delta=(\overline{d^2} - \overline{d}^2)^{1/2}/\overline{d}$, where the overline denotes an average over the distribution $f(d)$. Here we choose  $\delta = 0.23$ by imposing $d_{\rm min} / d_{\rm max} = 0.449$. The average diameter, $\overline{d}$, sets the unit of length. The soft-disk interactions are pairwise and described by an inverse power-law potential
\begin{eqnarray}
v_{ij}(r) &=& v_0 \left( \frac{d_{ij}}{r} \right)^{12} + c_0 + c_1 \left( \frac{r}{d_{ij}} \right)^2 + c_2 \left( \frac{r}{d_{ij}} \right)^4, \label{eq:soft_v} \\
d_{ij} &=& \frac{(d_i + d_j)}{2} (1-\epsilon |d_i - d_j|), \label{eq:non_additive}
\end{eqnarray}
where $v_0$ sets the unit of energy (and of temperature with the Boltzmann constant $k_\mathrm{B}\equiv 1$) and $\epsilon=0.2$  quantifies the degree of nonadditivity of particle diameters. We introduce $\epsilon>0$ in the model to suppress fractionation and thus enhance the glass-forming ability. The constants $c_0$, $c_1$ and $c_2$ enforce a vanishing potential and continuity of its first- and second-order derivatives at the cut-off distance $r_{\rm cut}=1.25 d_{ij}$.
We simulate a system with $N$ particles within a square cell of area $V=L^2$, where $L$ is the linear box length, under periodic boundary conditions, at a number density $\rho=N/V=1$.
We also compare the results with the corresponding results of the 3D system studied in Ref.~[\onlinecite{ozawa2018random}].

\subsection{Glass preparation}

Glass samples have been prepared by first equilibrating liquid configurations at a finite temperature, $T_{\rm ini}$ (which is sometimes referred to as the fictive temperature of the glass sample) and then performing a rapid quench to $T=0$, the temperature at which the samples are subsequently deformed. We prepare equilibrium configurations for the polydisperse disks using swap Monte-Carlo simulations~\cite{berthier2016equilibrium,ninarello2017models}. With probability $P_{\rm swap}=0.2$, we perform a swap move where we pick two particles at random and attempt to exchange their diameters, and with probability $1-P_{\rm swap}=0.8$, we perform conventional Monte-Carlo translational moves. To perform the quench from the obtained equilibrium configurations at $T_{\rm ini}$ down to zero temperature, we use the conjugate-gradient method~\cite{nocedal2006numerical}.

The preparation temperature $T_{\rm ini}$ then uniquely controls the stability of glass. We consider a wide range of preparation temperatures, from $T_{\rm ini} = 0.035$ to $T_{\rm ini}=0.200$. To better characterize this temperature span, we give some empirically determined representative temperatures of the model: Onset of slow dynamics~\cite{sastry1998signatures} takes place at $T_{\rm onset} \approx 0.23$, the dynamical mode-coupling crossover~\cite{gotze2008complex} at $T_{\rm mct} \approx 0.11$, and the estimated experimental glass transition temperature, obtained from extrapolation of the relaxation time~\cite{ninarello2017models}, at $T_{\rm g} \approx 0.068$. Note that these values are slightly different from the ones presented in Ref.~\cite{berthier2019zero} due to a small difference in the number density $\rho$. Our range of fictive temperature $T_{\rm ini}$ therefore covers from slightly below the onset temperature to significantly below the estimated experimental glass transition temperature.

\subsection{Mechanical loading}
 
We have performed strain-controlled athermal quasi-static shear (AQS) deformation using Lees-Edwards boundary conditions~\cite{maloney2006amorphous}.
The AQS shear method consists of a succession of tiny uniform shear deformation with $\Delta \gamma=10^{-4}$, followed by energy minimization via the conjugate-gradient method. The AQS deformation is performed along the $x$-direction up to the maximum strain $\gamma_{\rm max}=0.2$. Note that during the AQS deformation, the system is always located in a potential energy minimum (except of course during the transient conjugate-gradient minimization), i.e., it stays at $T=0$.

To obtain the averaged values of the various observables, $\langle (\cdots) \rangle$, in the simulations, we average over 800, 700, 400, 200, 200, 200, 200, and 100 samples for $N=1000$, $2000$, $4000$, $8000$, $16000$, $32000$, $64000$, and $128000$, respectively. 
For the lowest $T_{\rm ini}=0.035$, we average over $400$ samples for $N=8000$, $16000$, $32000$, and $64000$ systems.

\subsection{Nonaffine displacement}.

We consider the local nonaffine displacement of a given particle relative to its nearest neighbor particles, $D_{\rm min}^2$~\cite{falk1998dynamics}. $D_{\rm min}^2$ is always measured between the origin ($\gamma=0$) and a given strain $\gamma$. We define nearest neighbors by using the cut-off radius of the interaction range, $R_{\rm cut}=3.0 \overline{d}$. We determine the nearest neighbors of a particle from the configuration at $\gamma=0$.

\section{Results}
\label{sec:results}

\subsection{Nonequilibrium discontinuous yielding transition}
\label{sec:discontinuous_transition}

\begin{figure}
\includegraphics[width=0.49\columnwidth]{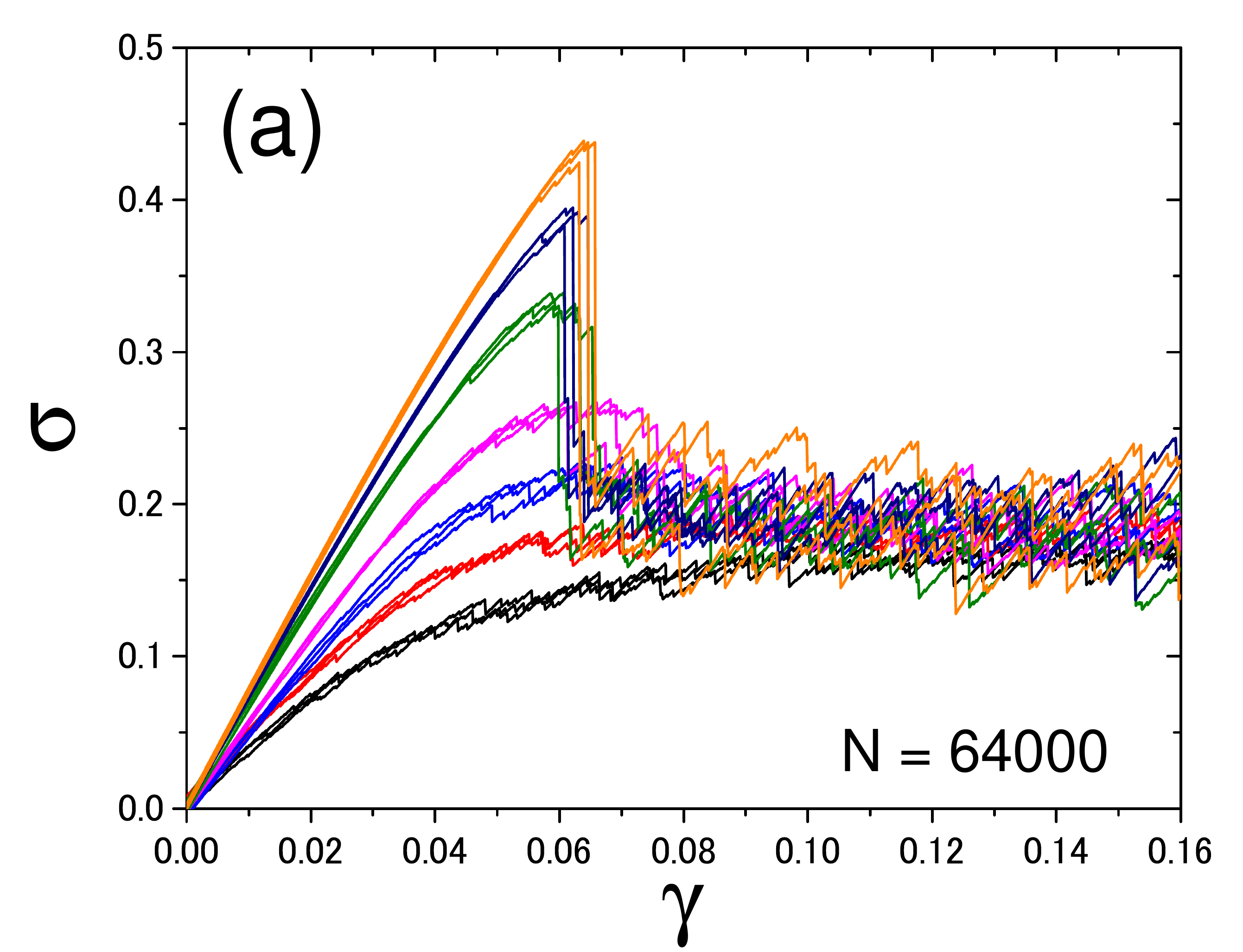}
\includegraphics[width=0.49\columnwidth]{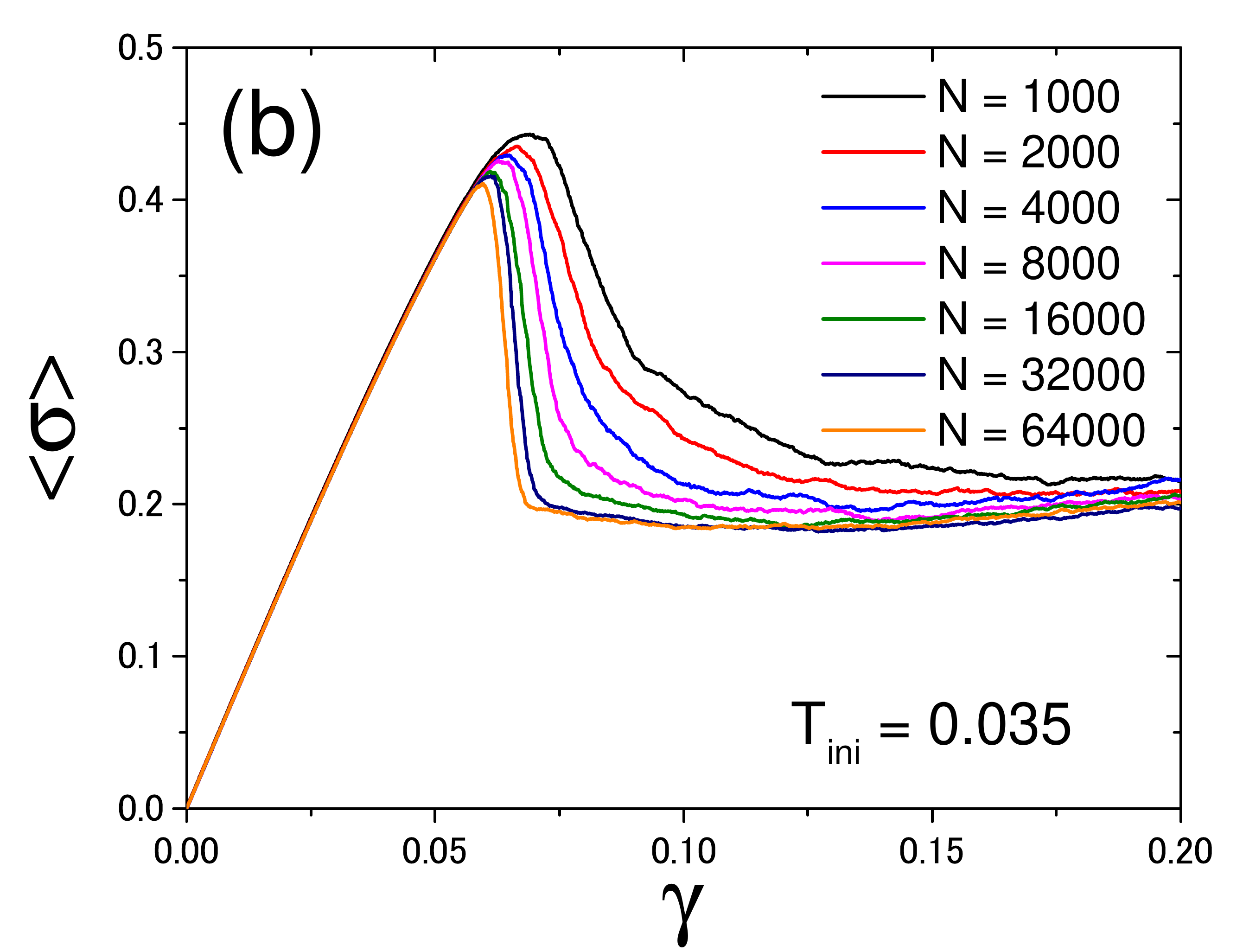}
\includegraphics[width=0.49\columnwidth]{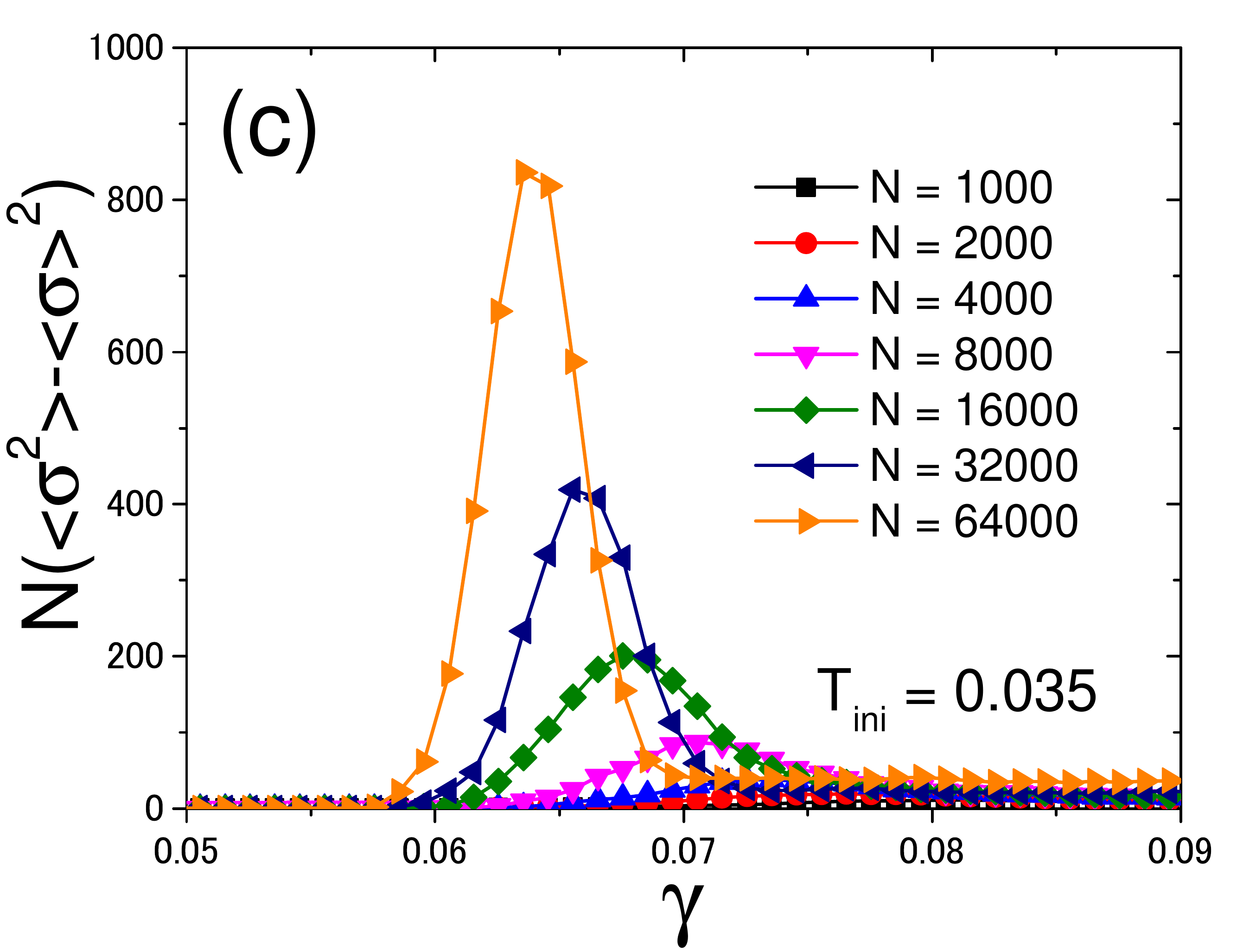}
\includegraphics[width=0.49\columnwidth]{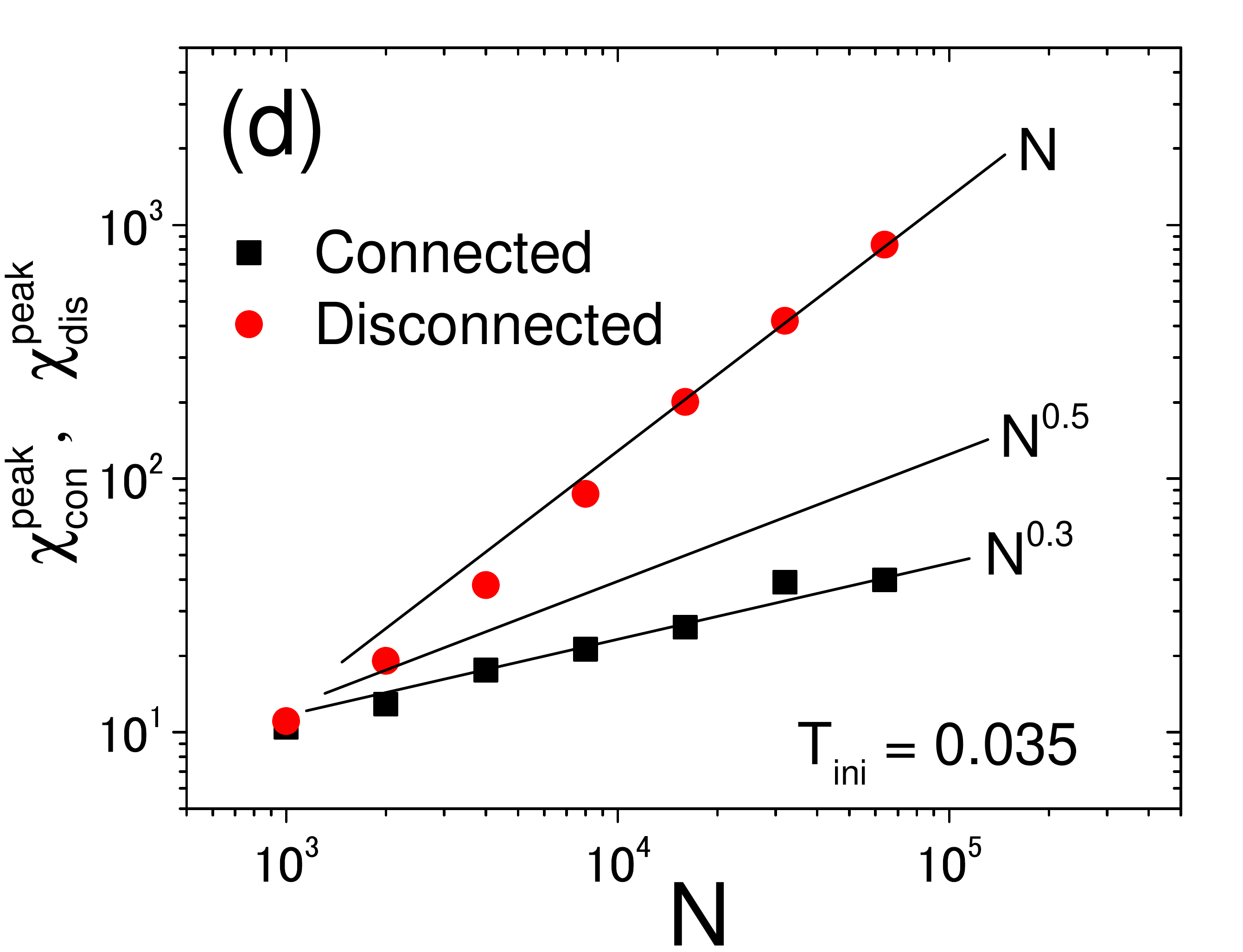}
\caption{
(a): Stress-strain curves for several typical samples of size $N=64000$ characterized for a wide span of preparation temperature $T_{\rm ini}$. From top to bottom; $T_{\rm ini}=0.035$, $0.050$, $0.070$, $0.100$, $0.120$, $0.150$, and $0.200$. For each $T_{\rm ini}$, three independent samples are shown.
(b): Averaged stress-strain curves at $T_{\rm ini}=0.035$ for several values of $N$.
(c): Disconnected susceptibility $\chi_{\rm dis}$ at $T_{\rm ini}=0.035$ for several values of $N$.
(d): Peak values of the connected and disconnected susceptibilities, $\chi_{\rm con}^{\rm peak}$ and $\chi_{\rm dis}^{\rm peak}$, at $T_{\rm ini}=0.035$.
The straight lines are the predicted scaling behaviors (see Appendix~\ref{sec:susceptibilities}).
}
\label{fig:FIG1}
\end{figure}  

Figure~\ref{fig:FIG1}(a) shows the stress-strain curves of typical samples for several values of $T_{\rm ini}$. The curves show different types of behavior depending on the initial stability: monotonic crossover for poorly annealed samples ($T_{\rm ini}=0.150-0.200$), mild stress overshoot for ordinary computer glass samples ($T_{\rm ini}=0.100-0.120$), and a sharp discontinuous stress drop for very stable samples ($T_{\rm ini}=0.035-0.070$). This plot is qualitatively similar to that found in 3D~\cite{ozawa2018random}: ductile yielding is observed for higher $T_{\rm ini}$ and appears to continuously transform into brittle yielding below $T_{\rm ini} \approx 0.10-0.12$. 

We first give evidence through finite-size scaling analysis that brittle yielding persists in 2D as a nonequilibrium first-order (or discontinuous) transition. For the most stable glass considered ($T_{\rm ini}=0.035$), we show in Figs.~\ref{fig:FIG1}(b) and (c) the stress-strain curves after averaging over many samples and the so-called ``disconnected'' susceptibility~\cite{nattermann1998theory}, 
\begin{equation}
\chi_{\rm dis}=N(\langle \sigma^2 \rangle - \langle \sigma \rangle^2),
\end{equation}
where $\langle \cdots \rangle$ denotes an average over samples. As $N$ is increased, the slope of the drop following the stress overshoot becomes steeper, suggesting that the averaged stress-strain curve shows a discontinuous jump as $N \to \infty$. 
As shown below, this is due to the sudden appearance of shear bands~\footnote{The ($D-1$) displacement along the shear band is accompanied by a change in the elastic strain energy in the bulk of the system that scales extensively with system size and explains the first-order character of the transition marked by a discontinuous stress drop of $\mathcal{O}(1)$.}. Concomitantly, the disconnected susceptibility $\chi_{\rm dis}$ grows with $N$. We plot its peak values as well as that of the so-called ``connected'' susceptibility~\cite{nattermann1998theory},
\begin{equation}
\chi_{\rm con} = - \frac{d \langle \sigma \rangle}{d \gamma}\,,
\end{equation}
in Fig.~\ref{fig:FIG1}(d). We find that both susceptibilities increase with $N$, $\chi_{\rm dis}^{\rm peak} \propto N$ and  $\chi_{\rm con}^{\rm peak} \propto N^{0.3}$, which is a signature of a nonequilibrium first-order (i.e., discontinuous) transition. The stronger divergence of $\chi_{\rm dis}^{\rm peak}$ indicates the predominant role of disorder fluctuations, as generically found in the RFIM.

The observed finite-size scaling of $\chi_{\rm dis}^{\rm peak}$ is the same in 2D and 3D; it reflects the discontinuous nature of the transition where sample-to-sample stress fluctuations at a fixed yield strain are of $\mathcal{O}(1)$. On the other hand, the scaling of $\chi_{\rm con}^{\rm peak}$ is different from 3D, for which we found $\chi_{\rm con}^{\rm peak} \propto N^{0.5}$~\cite{ozawa2018random}. The dominant effect explaining this difference comes from the scaling of the width of the distribution of $\gamma_{Y}$ at which the largest stress drop takes place. Sample-to-sample fluctuations seem to lead to a standard $N^{-0.5}$ behavior in 3D but to a broader distribution with a width decaying only as $N^{-0.3}$ in 2D (see Appendix~\ref{sec:susceptibilities}). Within a RFIM perspective~\cite{tarjus2008nonperturbative}, this entails that the variance of the effective random field at the transition scales with the linear system size $L=N^{1/D}$ as
\begin{equation}
\Delta_L= \frac{\chi_{\rm dis}^{\rm peak}}{(\chi_{\rm con}^{\rm peak})^2} \sim L^{\rho}
\end{equation}
with $\rho\approx 0.8$ in 2D and $\rho\approx 0$ in 3D. This in turn implies that the random field at yielding has long-range correlations decaying with distance as $r^{-(D-\rho)}$ with $\rho>0$~\cite{bray1986long,baczyk2013dimensional} in 2D, a feature that seems absent in 3D yielding. Note that the properties of the effective random field at the yielding transition result from a highly nontrivial combination of the disorder associated with the initial configurations and the evolution under deformation~\cite{rossi19}. This combination may vary with space dimension, albeit at present in a way that is not theoretically predicted.

\begin{figure}
\includegraphics[width=0.49\columnwidth]{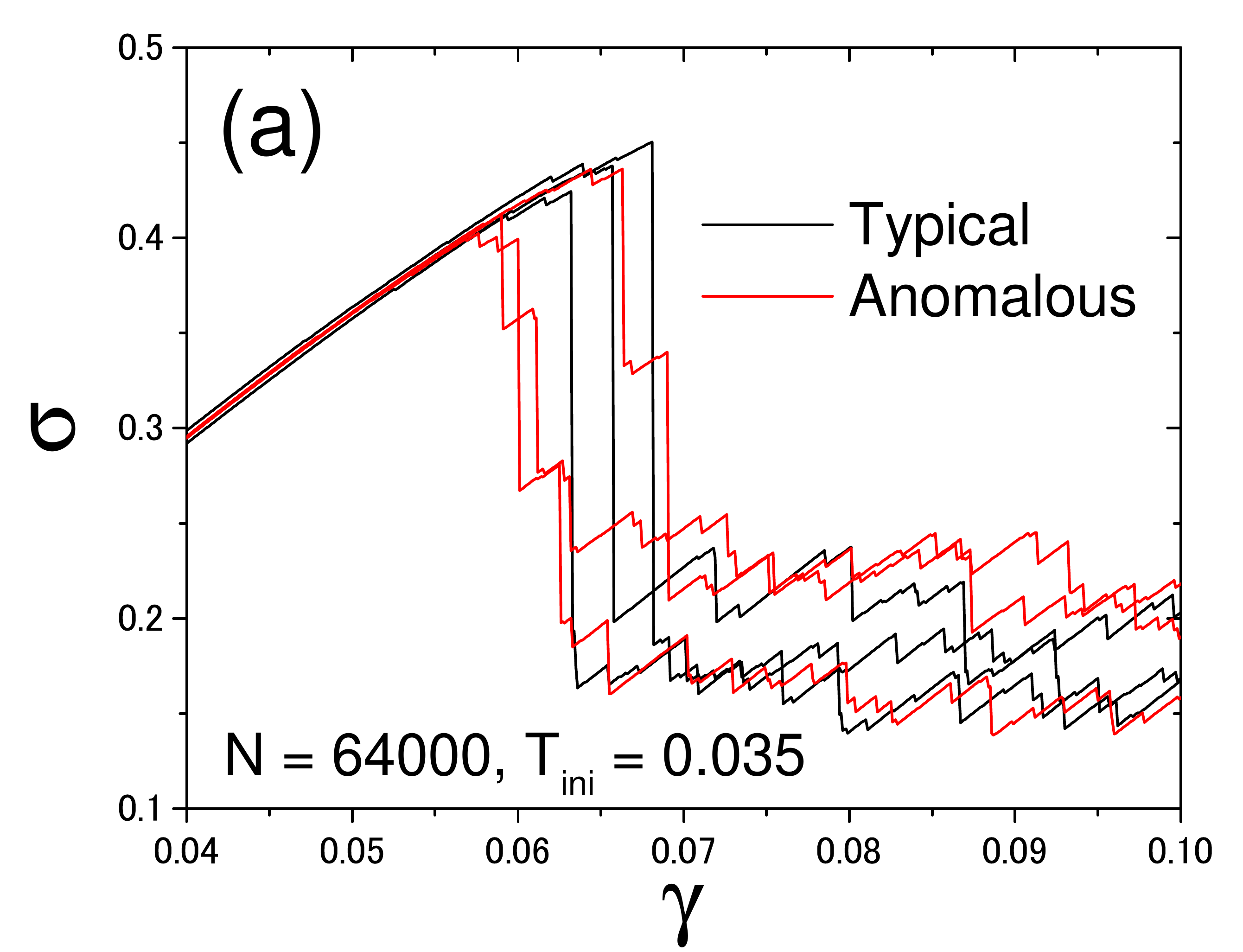}
\includegraphics[width=0.49\columnwidth]{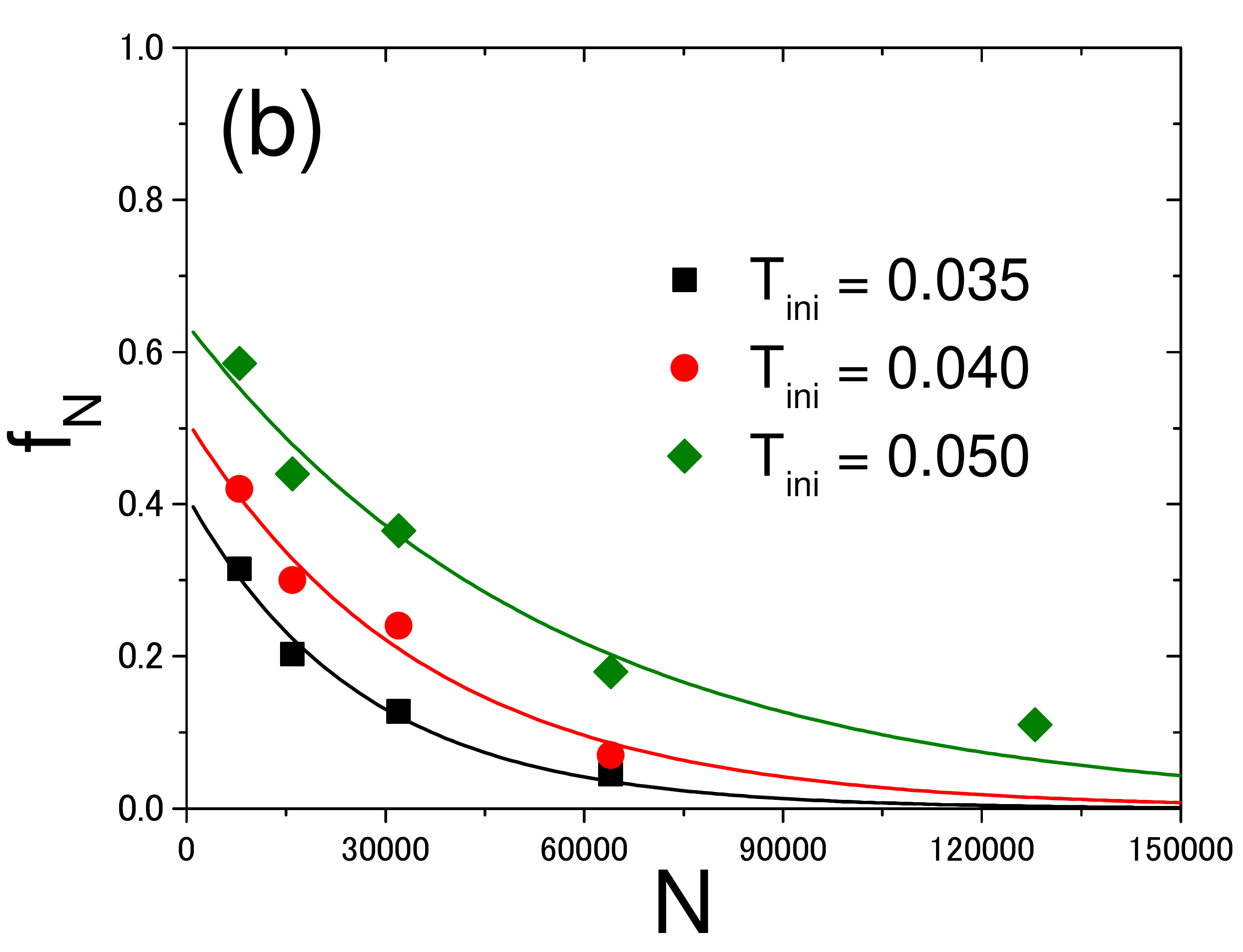}
\includegraphics[width=0.49\columnwidth]{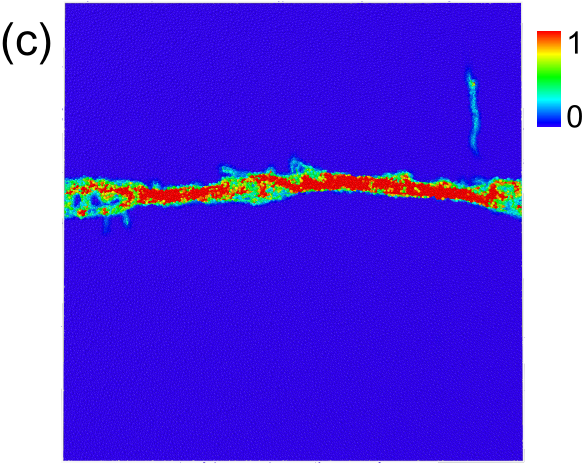}
\includegraphics[width=0.49\columnwidth]{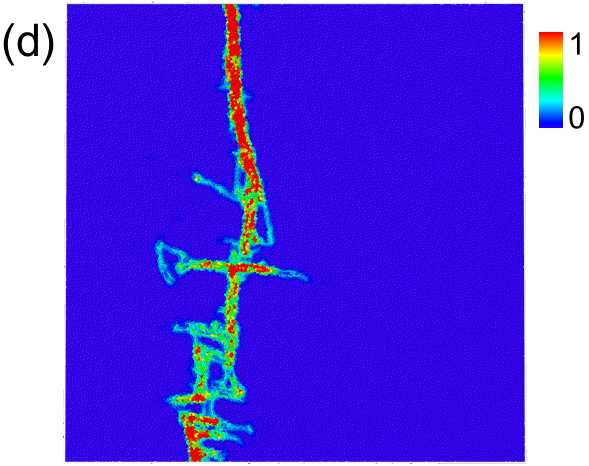}
\caption{
(a): Stress-strain curves for six samples with $N=64000$, $T_{\rm ini}=0.035$. The three black curves present typical samples showing a large stress drop. The three other red curves present rare samples showing multiple stress drops (which we call ``anomalous'' samples). 
(b): Fraction $f_N$ of anomalous samples as a function of $N$ for stable glasses. The solid curves are exponential fits of the data.
(c, d): Representative snapshots for typical (c) and anomalous samples (d) at $\gamma=0.07$. The color bar corresponds to the value of the nonaffine displacement, $D_{\rm min}^2$.
}
\label{fig:bad_samples}
\end{figure}

\subsection{Anomalous samples}
\label{sec:anomalous}

To further illustrate the strong sample-to-sample fluctuations present in 2D, even in the case of very stable glasses, we show in Fig.~\ref{fig:bad_samples}(a) a zoomed-in plot of the stress-strain curves for a few chosen samples. One can see that in addition to typical samples that display a single sharp, large stress drop (see also Fig.~\ref{fig:FIG1}(a)) there are samples that yield through multiple stress drops. These samples, which we refer to as ``anomalous'', display shear bands at yielding that tend to strongly wander and splinter in space (see Fig.~\ref{fig:bad_samples}(d)), whereas typical samples yield via the appearance of a well-defined system-spanning shear band (see Fig.~\ref{fig:bad_samples}(c))~\footnote{Note that due to the choice of periodic boundary conditions the direction of the shear band is either horizontal or vertical, with about equal probability for one or the other~\cite{kapteijns2019fast}. We have checked that the properties of horizontal and vertical shear bands are virtually the same.}. Anomalous samples lead to very large fluctuations but their fraction, $f_N$, decreases as $N$ increases. To quantify this effect, we have identified these samples from individual stress-strain curves by using the conditions $\Delta \sigma_{\rm max} \leq 0.15$, where $\Delta \sigma_{\rm max}$ is the maximum stress drop observed in the strain window $\gamma \in [0, \gamma_{\rm max}]$ for each sample (see Appendix~\ref{sec:appendix_anomalous} for details). In Fig.~\ref{fig:bad_samples}(b) we observe that $f_N$ decreases with $N$, in an apparently exponential manner, and appears to vanish as  $N \to \infty$: hence the terminology ``anomalous'' versus ``typical''. (Note that we have checked that the finite-size scaling of $\chi_{\rm dis}^{\rm peak}$ and $\chi_{\rm con}^{\rm peak}$ in FIG.~\ref{fig:FIG1}(d) hardly changes when we remove the anomalous samples from the computation; this shows  that the observed value $\approx 0.3$ for the scaling exponent of $\chi_{\rm con}^{\rm peak}$ is not due to the presence of the anomalous samples.) Repeating the same analysis for 3D glass samples of a similar stability, we find that $f_N$ virtually vanishes above $N=12000$ (not shown), which indicates much weaker finite-size effects than in 2D. 

\begin{figure}
\includegraphics[width=0.4\columnwidth]{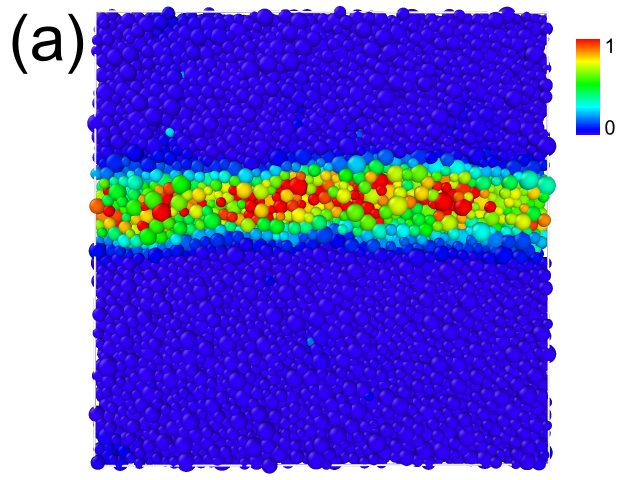}
\includegraphics[width=0.4\columnwidth]{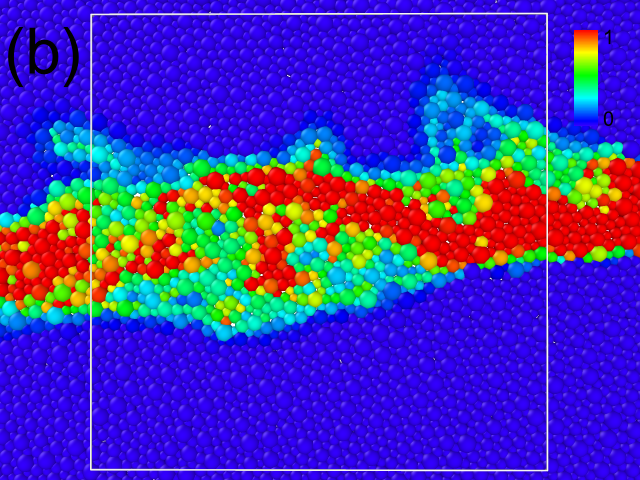}
\includegraphics[width=0.95\columnwidth]{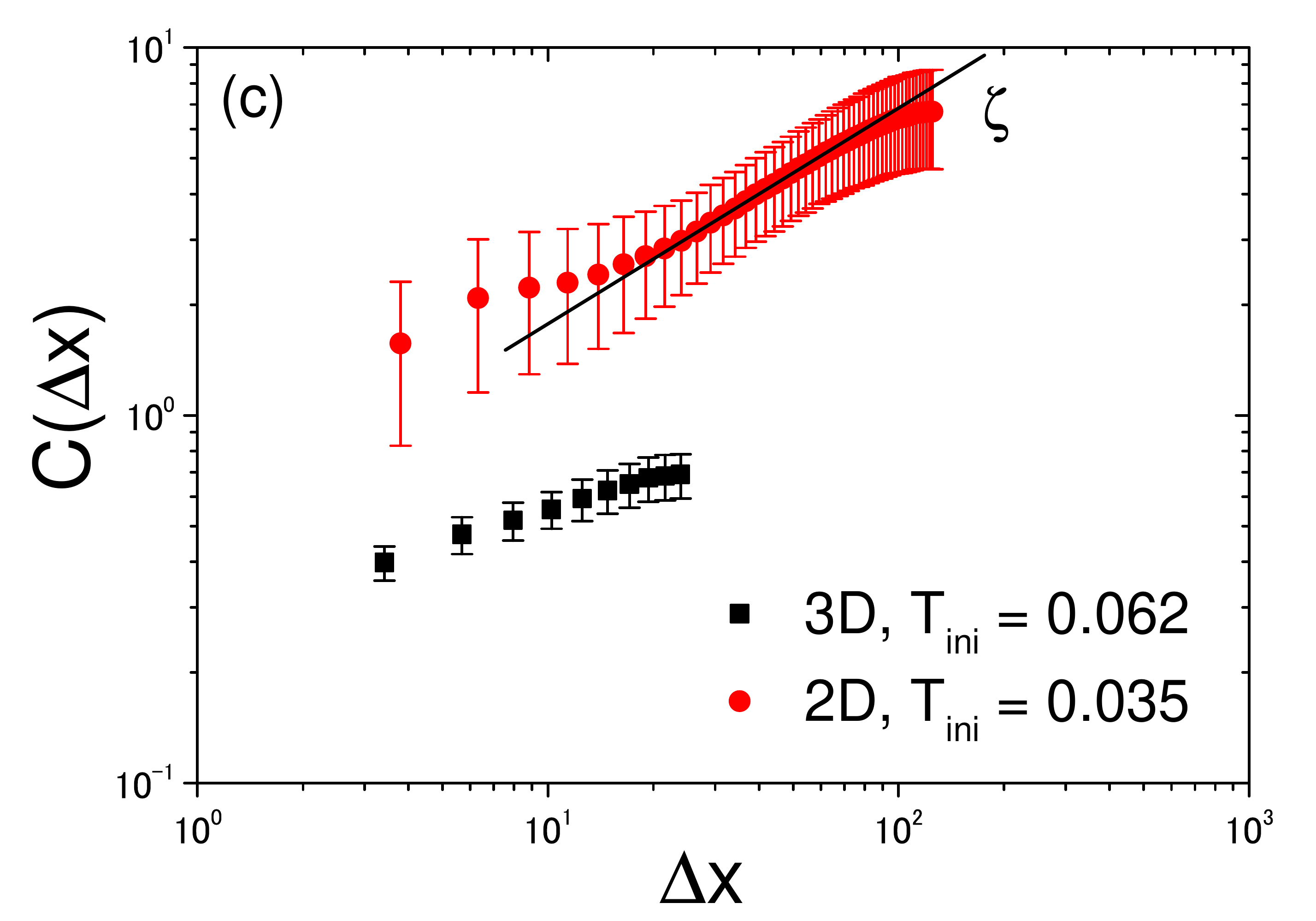}
\caption{
(a, b): Snapshot of shear bands in typical glass samples of similar stability in 3D ($N=96000$, $T_{\rm ini}=0.062$ at $\gamma=0.13$) (a) and in 2D (a zoomed-in plot of Fig.~\ref{fig:bad_samples}(c); $N=64000$, $T_{\rm ini}=0.035$ at $\gamma=0.07$) (b). The color bars correspond to the value of the nonaffine displacement $D_{\rm min}^2$. The white square region in (b) has the same area as the surface of the 3D simulation box.
(c): Height-height correlation function $C(\Delta x)$ for 2D and 3D. The error bars correspond to the standard deviation from sample-to-sample fluctuations. 
The straight line corresponds to $\zeta \approx 0.59$.
}
\label{fig:roughness}
\end{figure}

\subsection{Rough spacially wandering shear bands}
\label{sec:rough}

Next, we investigate the spatial characteristics of the shear bands. We carefully analyze the wandering of the shear bands  in space, comparing 2D and 3D at the same lengthscale and for similar glass stability~\footnote{Similar glass stability in a sence that the ratio between the peak and steady state stress values are similar~\cite{kapteijns2019fast}.}. In Figs.~\ref{fig:roughness}(a) and (b), we show snapshots right after yielding for 3D ($\gamma=0.13$) and 2D ($\gamma=0.07$), respectively. The 2D shear band appears thicker than the 3D one, and a quantitative comparison is provided in the SI. More importantly, the shear band seems to wander more or, said otherwise, to be ``rougher'' in 2D. This roughness can be quantified by the height-height correlation function~\cite{bouchaud1997scaling,ponson2006two,alava2006statistical,bouchbinder2006fracture}: 
\begin{equation}
\label{eq_height}
C(\Delta x) = \langle (Y_{\rm com}(x+ \Delta x)-Y_{\rm com}(x))^2 \rangle_{x}^{1/2},
\end{equation}
where $Y_{\rm com}(x)$ is the average height of the shear band and $\langle \cdots \rangle_{x}$ denotes a spatial average. 
Note that we exclude the anomalous samples from the analysis, because defining the shear band interfaces is hard and often ambiguous in anomalous samples, e.g., when a shear band forms a closed-loop structure. For the 3D case we use an expression analogous to Eq.~(\ref{eq_height}) which also takes into account the average in the additional $z$ coordinate (see Appendix~\ref{sec:appendix_rough} for a detailed explanation). A manifold is rough on large scales if the height-height correlation function scales with distance as $C(\Delta x) \propto \Delta x^{\zeta}$, where $\zeta>0$ is the roughness exponent. Therefore, the log-log plot of $C(\Delta x)$ versus the distance $\Delta x$ provides a way to assess the roughness of the shear bands. We show such a plot in Fig.~\ref{fig:roughness}(c). 
The data in 3D show no convincing effect over the (limited) covered range but the results in 2D point to a nontrivial intermediate regime (limited at the longest lengthscales by a saturation due to the system size) where an effective roughness exponent $\zeta \approx 0.59$ can be observed~\cite{ponson2016statistical}. It is also clear that the overall magnitude of $C(\Delta x)$ in 2D is much larger than the one in 3D, and this difference is expected to grow even larger in larger samples. This again reflects the presence of larger spatial fluctuations in 2D.

\subsection{A critical point separates brittle and ductile yielding}
\label{sec:critical}

Because we have found strong evidence that yielding in 2D is a genuine nonequilibrium discontinuous transition for very stable glasses and that poorly annealed samples clearly show a continuous ductile behavior (see Fig.~\ref{fig:FIG1}(a)), it is tempting to look for signatures of a critical point separating brittle and ductile yielding as one varies the preparation temperature $T_{\rm ini}$ of the glass samples. In Ref.~\cite{ozawa2018random}, we showed that the difference between  the stress before ($\sigma_1$) and after ($\sigma_2$) the largest stress drop, $\Delta \sigma_{\rm max} = \sigma_1 - \sigma_2$,  plays the role of the order parameter distinguishing brittle from ductile yielding in 3D. In particular, we demonstrated that  the critical point at $T_{\rm ini,c}$ can be identified by the divergence of the variance of this order parameter, $N(\langle \Delta \sigma_{\rm max}^2 \rangle - \langle \Delta \sigma_{\rm max} \rangle^2)$.
In 2D however, as seen in Fig.~\ref{fig:FIG1}(a), strong fluctuations seem to also affect the plastic steady state, irrespective of the presence of a critical point. Even for typical samples, this blurs the determination of the largest stress drop when the latter becomes small as one approaches the putative critical point. As an operational procedure to remove this effect, we have therefore defined the order parameter as $\Delta \sigma_{\rm max} \equiv \sigma_1 - \langle \sigma_2 \rangle$. In this way the fluctuations of $\sigma_2$ that we tentatively attribute to the plastic steady-state regime are explicitly removed. When applied to 3D this new definition captures the critical point even more sharply without affecting the results. 

\begin{figure}
\includegraphics[width=0.95\columnwidth]{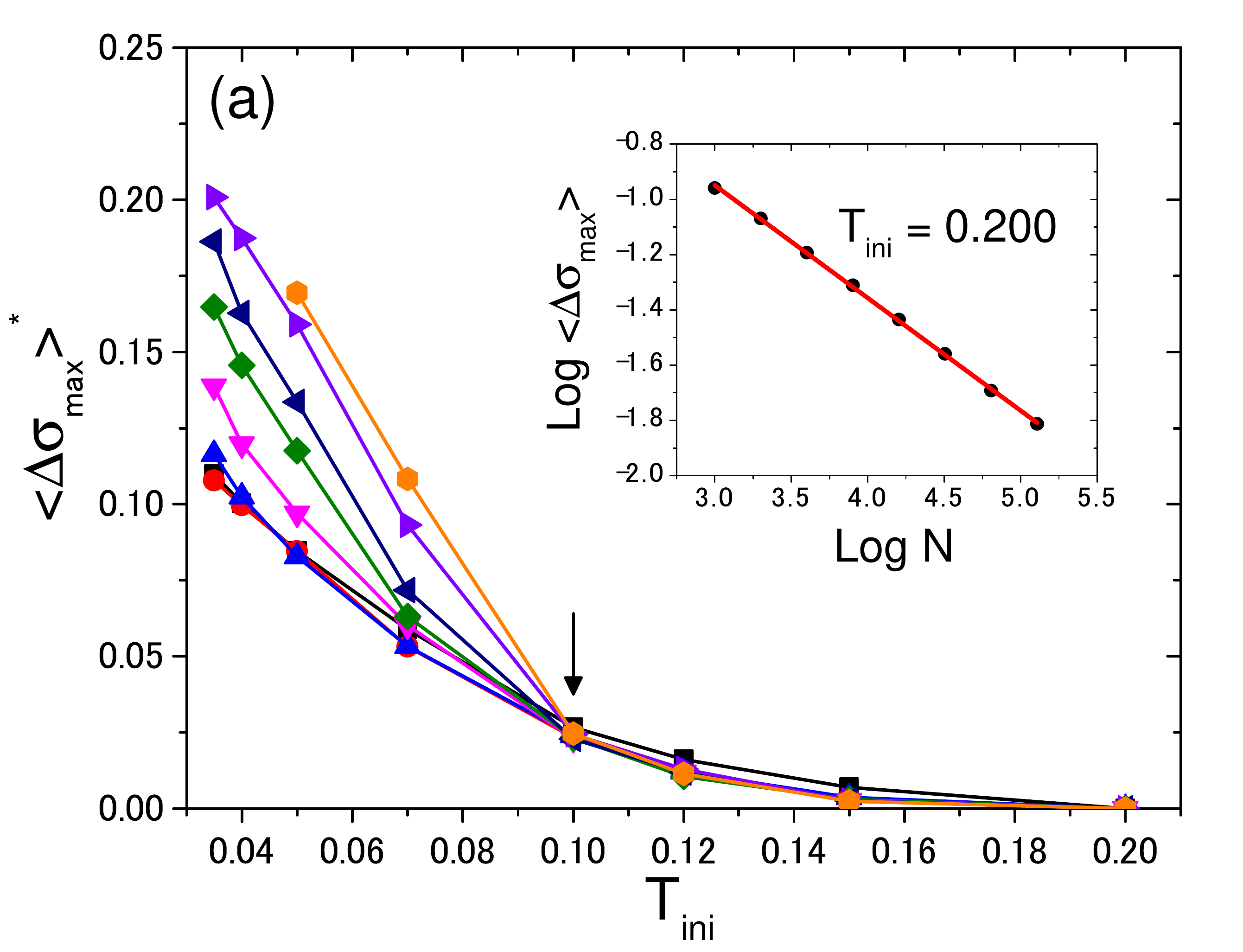}
\includegraphics[width=0.95\columnwidth]{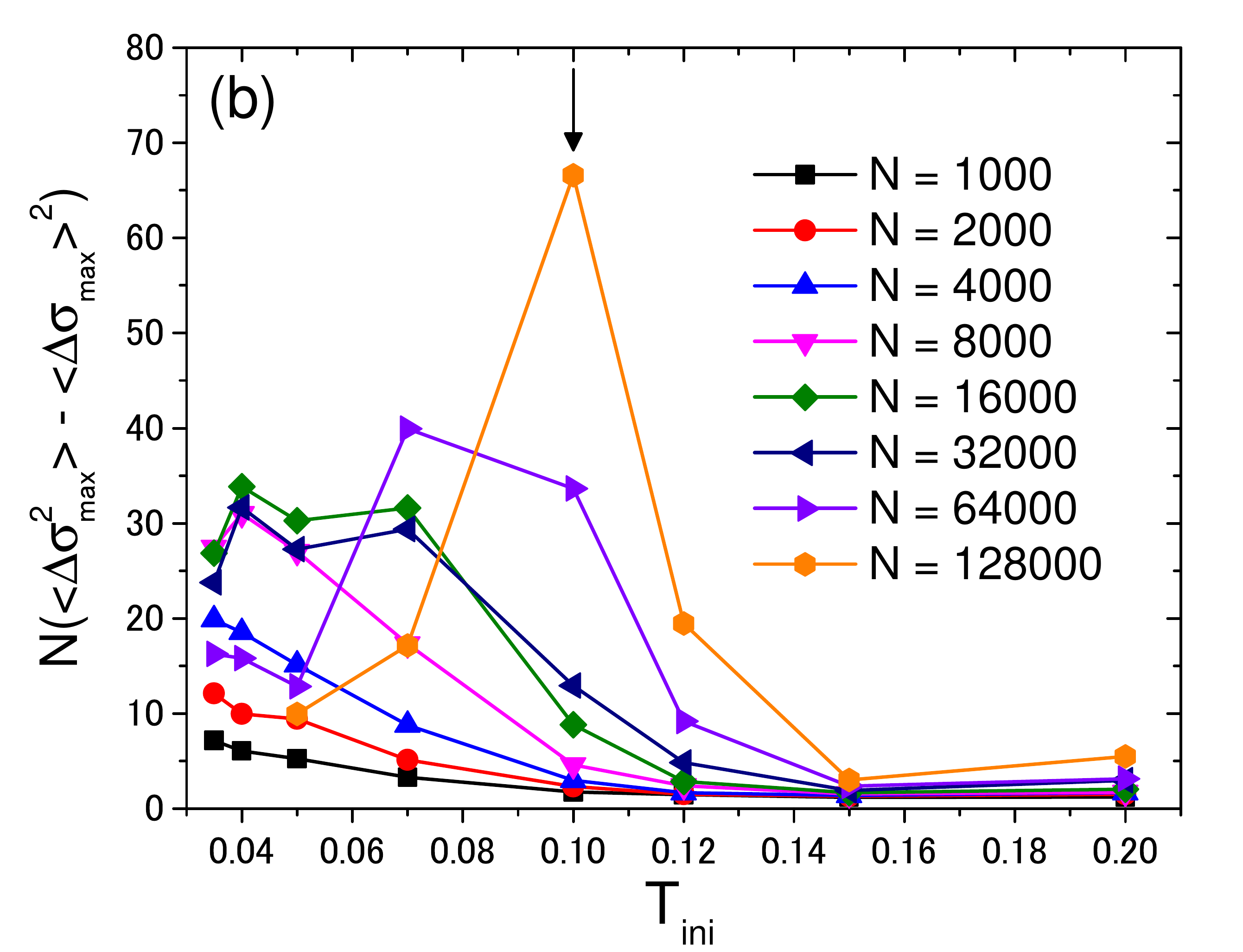}
\caption{
Mean (a) and variance (b) of the largest stress drop $\Delta \sigma_{\rm max}$ as a function of the glass preparation temperature $T_{\rm ini}$ for several system sizes $N$. In (a) we plot $\langle \Delta \sigma_{\rm max} \rangle - \langle \Delta \sigma_{\rm max} \rangle |_{T_{\rm ini}=0.2}$. We subtract the trivial high-temperature dependence that vanishes at large $N$, as shown in the inset where the data are fitted with a power law, $\langle \Delta \sigma_{\rm max} \rangle \propto N^{-0.41}$. The vertical arrows in (a) and (b) indicate the putative critical point associated with the brittle-to-ductile transition in 2D.}
\label{fig:critical}
\end{figure}

The mean value $\langle \Delta \sigma_{\rm max} \rangle$ is shown in  Fig.~\ref{fig:critical}(a) (we have removed a trivial offset at high $T_{\rm ini}$ which vanishes in the large-$N$ limit, see the inset): $\langle \Delta \sigma_{\rm max} \rangle$ appears rather flat at high $T_{\rm ini}$ and starts to grow and to develop a significant dependence on system size below $T_{\rm ini} \approx 0.1$, showing a similar trend as the 3D case. The variance of $\Delta \sigma_{\rm max}$ is displayed in Fig.~\ref{fig:critical}(b). It shows a peak that grows and shifts toward higher $T_{\rm ini}$ with $N$. The data is not sufficient to allow for a proper determination of critical exponents but give nonetheless support to the existence of a brittle-to-ductile {\it critical point} around $T_{\rm ini,c} \approx 0.1$. 

\section{Conclusions}
\label{sec:conclusions}

We have given strong numerical evidence that 2D yielding of very stable glasses under athermal  quasi-static shear remains a nonequilibrium first-order (discontinuous) transition that survives in the thermodynamic limit, with a dominance of the disorder-induced, i.e., sample-to-sample, fluctuations. Furthermore, the transition to ductile yielding is signalled by a critical point. The scenario found in 2D is therefore analogous to the one found in 3D,
but with stronger fluctuation effects. On the one hand, this suggests that the brittle-to-ductile transition as a function of sample preparation could be also experimentally observed in 2D or quasi-2D amorphous materials. On the other hand, it confirms that if indeed the effective theory describing the yielding transition is an athermally quasi-statically driven RFIM, the basic features of the model are necessarily modified by the presence of long-range anisotropic Eshelby-like interactions and, in 2D possibly, by long-range correlations in the effective random field. The long-range and quadrupolar nature of the elastic interactions accounts for the appearance of a shear band at the spinodal point marking the start of brittle yielding in stable glasses.
In this modified RFIM, the nature of the spinodal and its potential critical character~\cite{parisi2017shear} still need to be investigated.

\begin{acknowledgments}
We acknowledge support from the Simons Foundation 
(\#454933, L. Berthier, \#454935, G. Biroli). 
\end{acknowledgments}

\appendix

\section{Susceptibilities}
\label{sec:susceptibilities}

\subsection{Computation of the susceptibilities} 

\begin{figure}
\includegraphics[width=0.95\columnwidth]{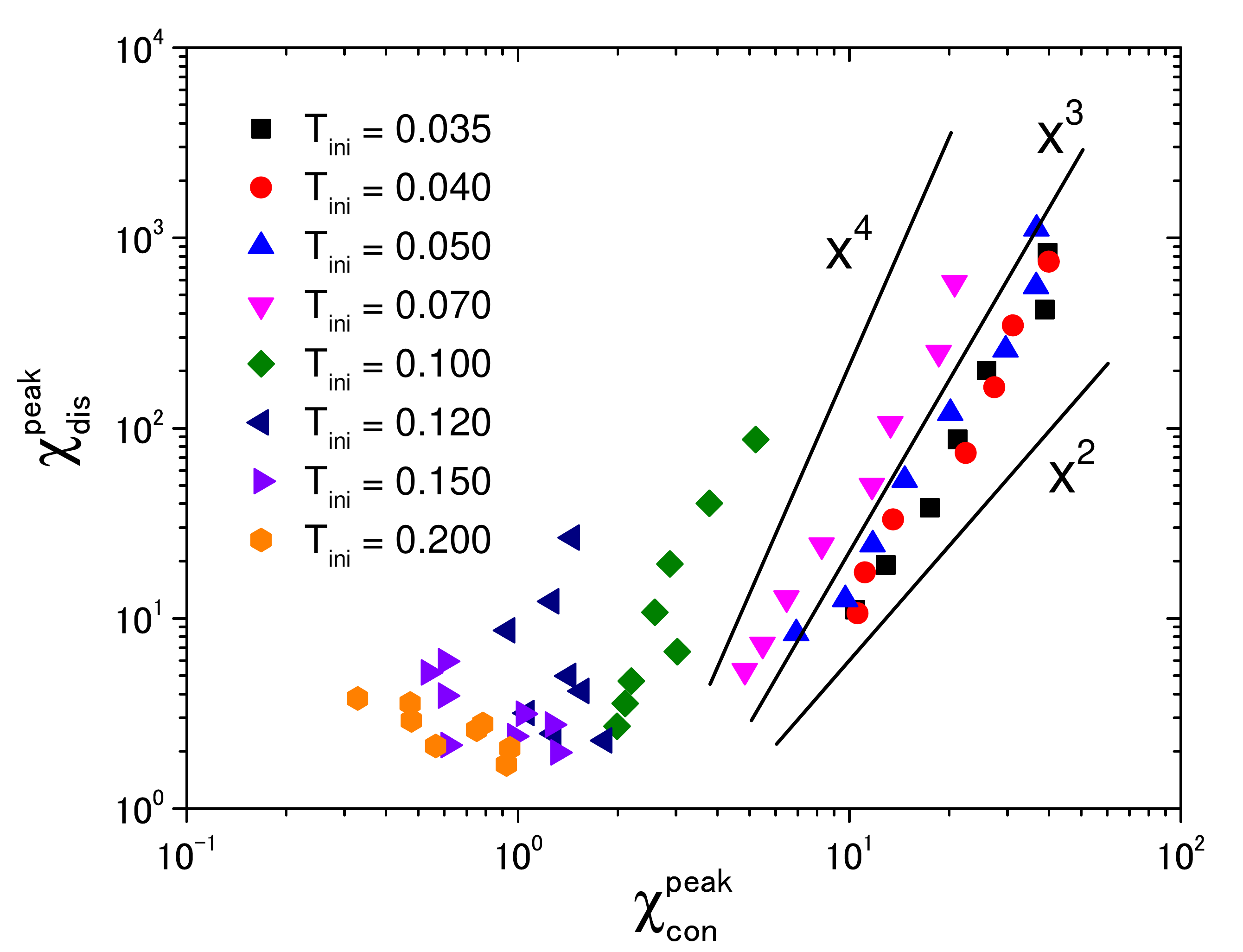}
\caption{
Parametric (log-log) plot of the peak values of the disconnected and connected susceptibilities for all system sizes and preparation temperatures. For $T_{\rm ini} \lesssim 0.10$ the logarithm of the disconnected susceptibility grows with the logarithm of the connected one.}
\label{fig:con_vs_dis}
\end{figure}

To numerically  compute the susceptibilities, we perform a smoothing procedure by averaging over 10 adjacent data points, as described in Ref.~[\onlinecite{ozawa2018random}]. Figure~\ref{fig:con_vs_dis} shows the parametric plot of the logarithms of the peak values of $\chi_{\rm dis}$ and $\chi_{\rm con}$. The data points roughly follow in a straight line in the putative brittle yielding phase ($T_{\rm ini} \lesssim 0.10$), with a slope of about $3$.

\subsection{Finite-size scaling of the susceptibilities for a discontinuous transition in the presence of disorder} 

\begin{figure*}
\includegraphics[width=0.66\columnwidth]{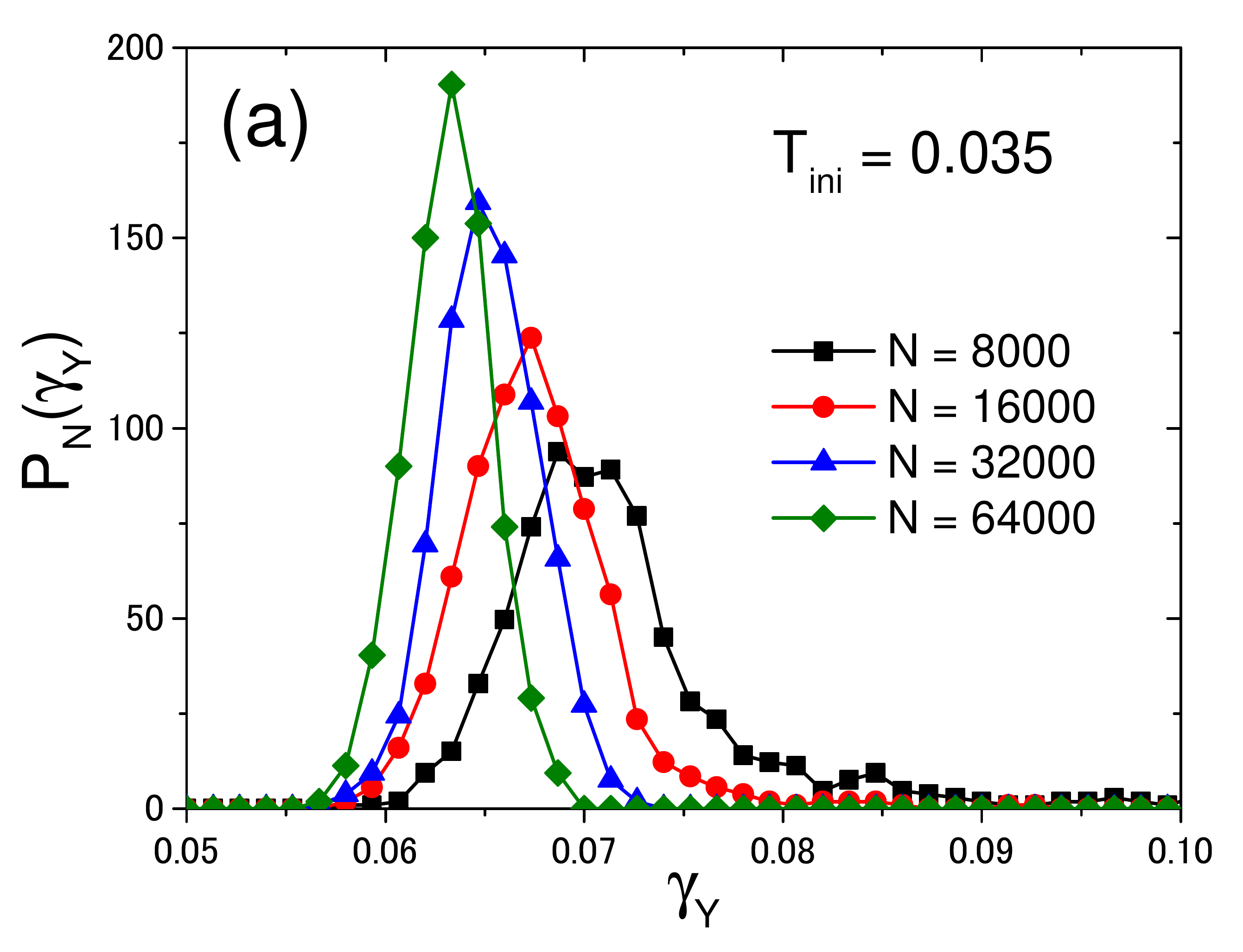}
\includegraphics[width=0.66\columnwidth]{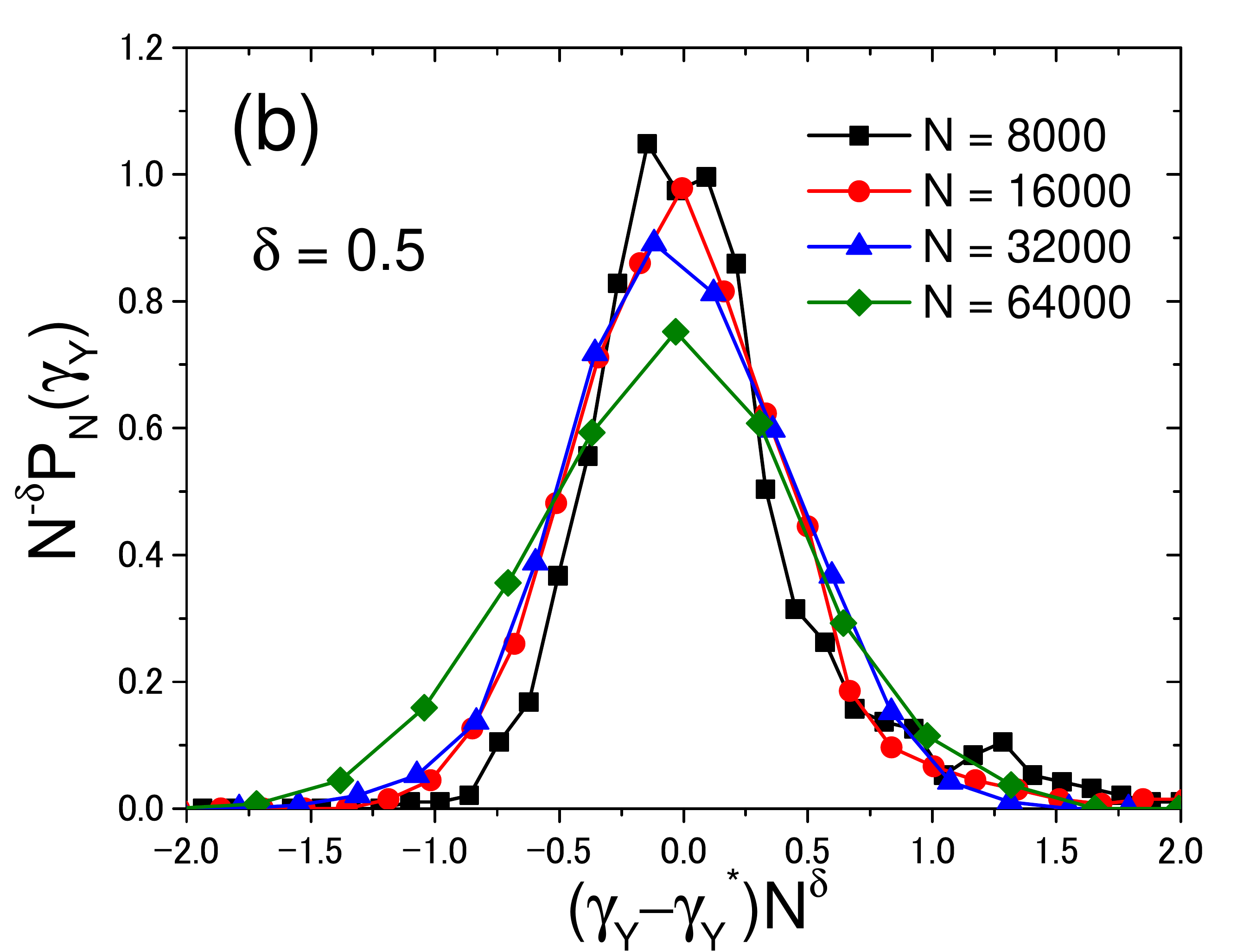}
\includegraphics[width=0.66\columnwidth]{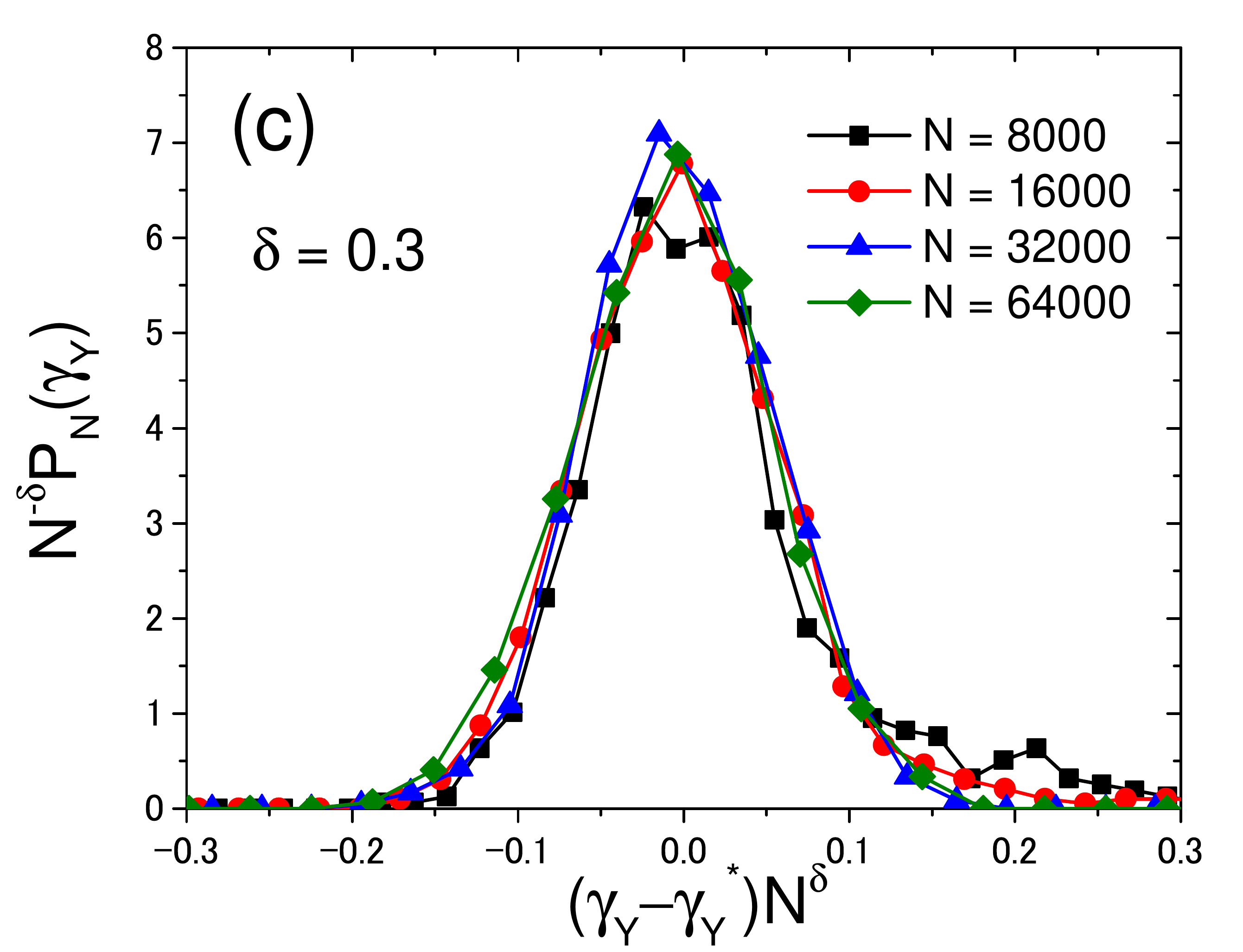}
\caption{(a): Probability distribution of $\gamma_{\rm Y}$ determined by the position of the largest stress drop, $\Delta \sigma_{\rm max}$.
(b, c): Scaling analysis assuming Eq.~(\ref{eq:probab_yieldstrain}) with $\delta=0.5$ (a) and $\delta=0.3$ (b).}
\label{fig:modeling}
\end{figure*}

We consider the case of a very stable glass (appropriate, e.g., for $T_{\rm ini}=0.035$), for which each typical finite-size sample yields through a discontinuous stress drop, $\Delta\sigma_{\rm max}$, of $\mathcal{O}(1)$ at a strain value $\gamma_{\rm Y}$. Both the size of the stress drop and the yield strain are sample dependent. It is easily realized that provided the mean value of $\langle \Delta\sigma_{\rm max} \rangle$ is strictly positive and of $\mathcal{O}(1)$, the fluctuations of $\Delta\sigma_{\rm max}$ lead only to subdominant contributions to the finite-size scaling of the susceptibilities, at least well below $T_{\rm ini, c}$. The fluctuations of the yield strain on the other hand are crucial. They are likely to regress with the system size $N$ but possibly in a nontrivial fashion. We assume that the values of $\gamma_{\rm Y}$ are distributed around the peak position, $\gamma_{\rm Y}^*$, of the distribution according to some probability function scaling with $N$ as
\begin{equation}
\label{eq:probab_yieldstrain}
P_N(\gamma_{\rm Y})\sim N^{\delta} \mathcal{P} \left( (\gamma_{\rm Y}- \gamma_{\rm Y}^*)N^\delta \right),
\end{equation}
with $\delta>0$ and $\mathcal{P}(0)>0$. We have computed this distribution of $\gamma_{\rm Y}$ for 2D samples prepared at $T_{\rm ini}=0.035$ and the result is shown in Fig.~\ref{fig:modeling}(a). 
We then perform exercises of scaling collapse assuming Eq.~(\ref{eq:probab_yieldstrain}) with various $\delta$ in Figs.~\ref{fig:modeling}(b,c).
We find that the conventional value, $\delta=0.5$~\cite{procaccia2017mechanical}, does not work, while a smaller value, $\delta=0.3$, provides a good scaling collapse.
In the following, we will relate the obtained value, $\delta=0.3$, with the scaling of the susceptibilities.

In the vicinity of yielding, one may describe the stress in each sample as given by 
\begin{equation}
\label{eq:stress-strain}
\sigma(\gamma)\approx \sigma_2+\Delta\sigma_{\rm max} \theta(\gamma_{\rm Y}-\gamma),
\end{equation}
where $\theta(x)$ is the Heaviside step function and $\sigma_2$ is the stress right after the stress drop. As discussed in the main text, this value may also fluctuate, but just as for the fluctuations of $\Delta\sigma_{\rm max}$ this leads to only subdominant corrections to the leading finite-size scaling in the regime where a strong discontinuous transition is present. We therefore assume from now on that neither $\Delta\sigma_{\rm max}$ nor $\sigma_2$ fluctuate from sample to sample. One then easily derives that the first two cumulants of $\sigma(\gamma)$ close to the yielding transition are expressed as
\begin{equation}
\label{eq:first-cumulant}
\langle \sigma(\gamma) \rangle =\sigma_2 +\Delta\sigma_{\rm max} \int_{\gamma}^{+\infty} d\gamma_{\rm Y} P_N(\gamma_{\rm Y}),
\end{equation}
and
\begin{equation}
\begin{aligned}
\label{eq:second-cumulant}
&\langle \sigma(\gamma)^2 \rangle-\langle \sigma(\gamma)\rangle^2=\\&(\Delta\sigma_{\rm max})^2 \int_{\gamma}^{+\infty} d\gamma_{\rm Y} P_N(\gamma_{\rm Y})\left(1-\int_{\gamma}^{+\infty} d\gamma_{\rm Y} P_N(\gamma_{\rm Y})\right).
\end{aligned}
\end{equation}

From the above expressions it is easy to derive the connected susceptibility, 
\begin{equation}
\chi_{\rm con}(\gamma)=-\frac{d \langle \sigma(\gamma) \rangle}{d\gamma}=\Delta\sigma_{\rm max}P_N(\gamma)
\end{equation}
and the disconnected one,
\begin{equation}
\begin{aligned}
&\chi_{\rm dis}(\gamma)=N \left( \langle \sigma(\gamma)^2 \rangle - \langle \sigma(\gamma) \rangle^2 \right)\\&=N(\Delta\sigma_{\rm max})^2 \int_{\gamma}^{+\infty} d\gamma_{\rm Y} P_N(\gamma_{\rm Y})\left( 1-\int_{\gamma}^{+\infty} d\gamma_{\rm Y} P_N(\gamma_{\rm Y})\right).
\end{aligned}
\end{equation}
By taking now into account the scaling form of the distribution $P_N(\gamma_{\rm Y})$ in Eq.~(\ref{eq:probab_yieldstrain}), one immediately obtains that the maximum of the susceptibilities scales as
\begin{equation}
\begin{aligned}
&\chi_{\rm con}^{\rm peak}\sim N^\delta \Delta\sigma_{\rm max},
\\&\chi_{\rm dis}^{\rm peak} \sim N (\Delta\sigma_{\rm max})^2.
\end{aligned}
\end{equation}
Putting $\delta=0.3$ obtained numerically by the scaling analysis in Fig.~\ref{fig:modeling}, 
we show a comparison between the above predictions and the directly determined $N$ dependence of $\chi_{\rm con}^{\rm peak}$ and $\chi_{\rm dis}^{\rm peak}$ in Fig. 1(d) of the main text. We find excellent agreements.

\begin{figure}
\includegraphics[width=0.48\columnwidth]{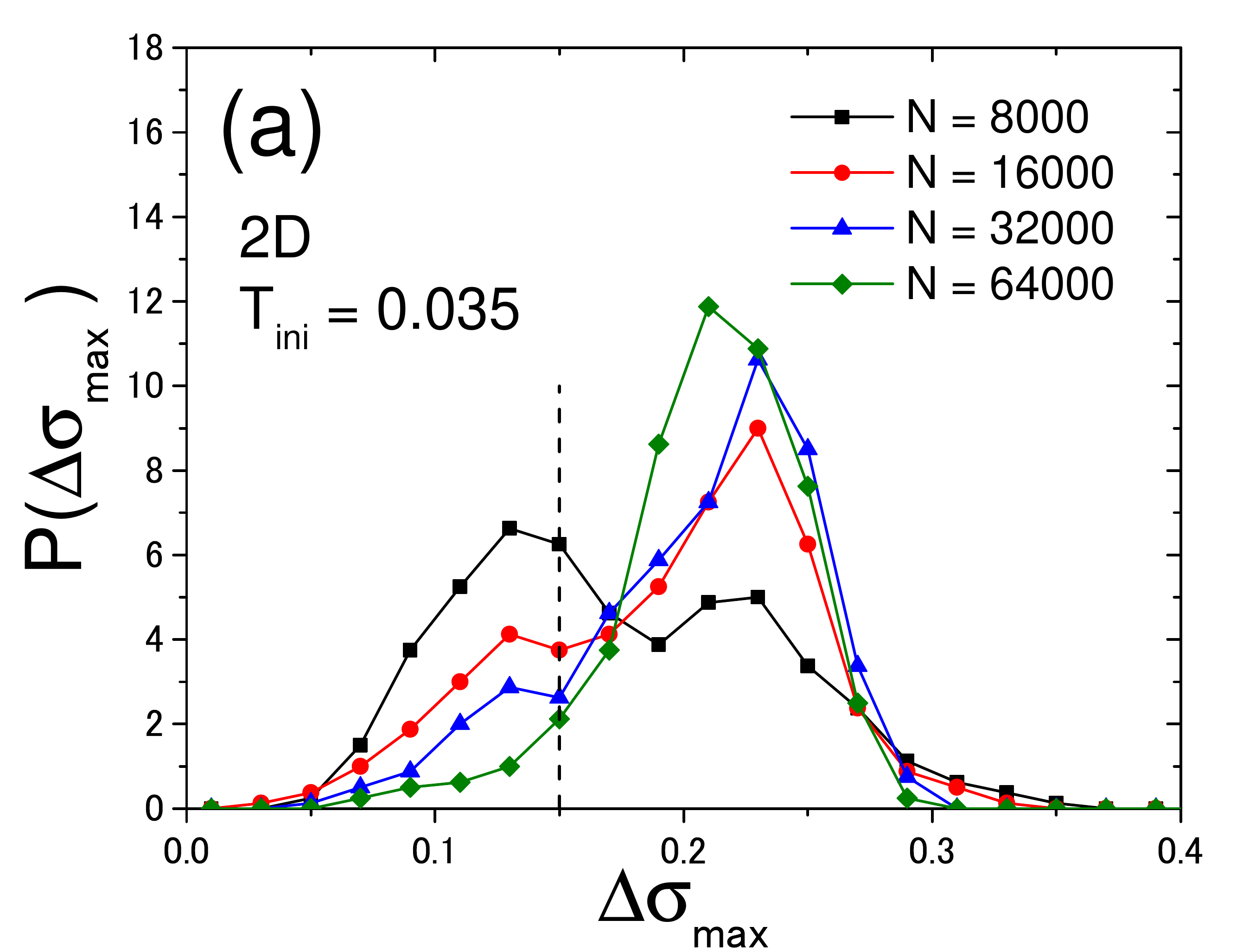}
\includegraphics[width=0.48\columnwidth]{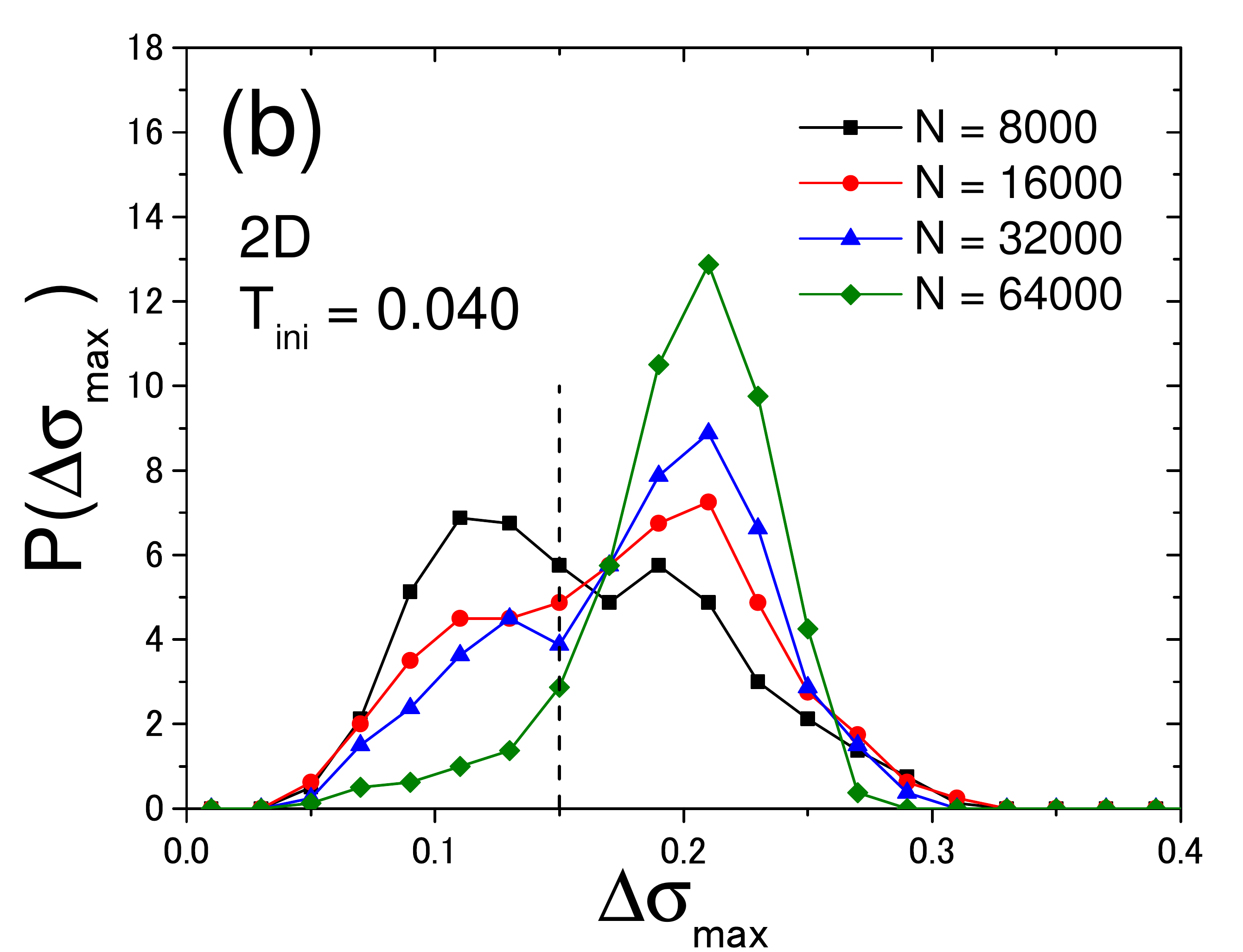}
\includegraphics[width=0.48\columnwidth]{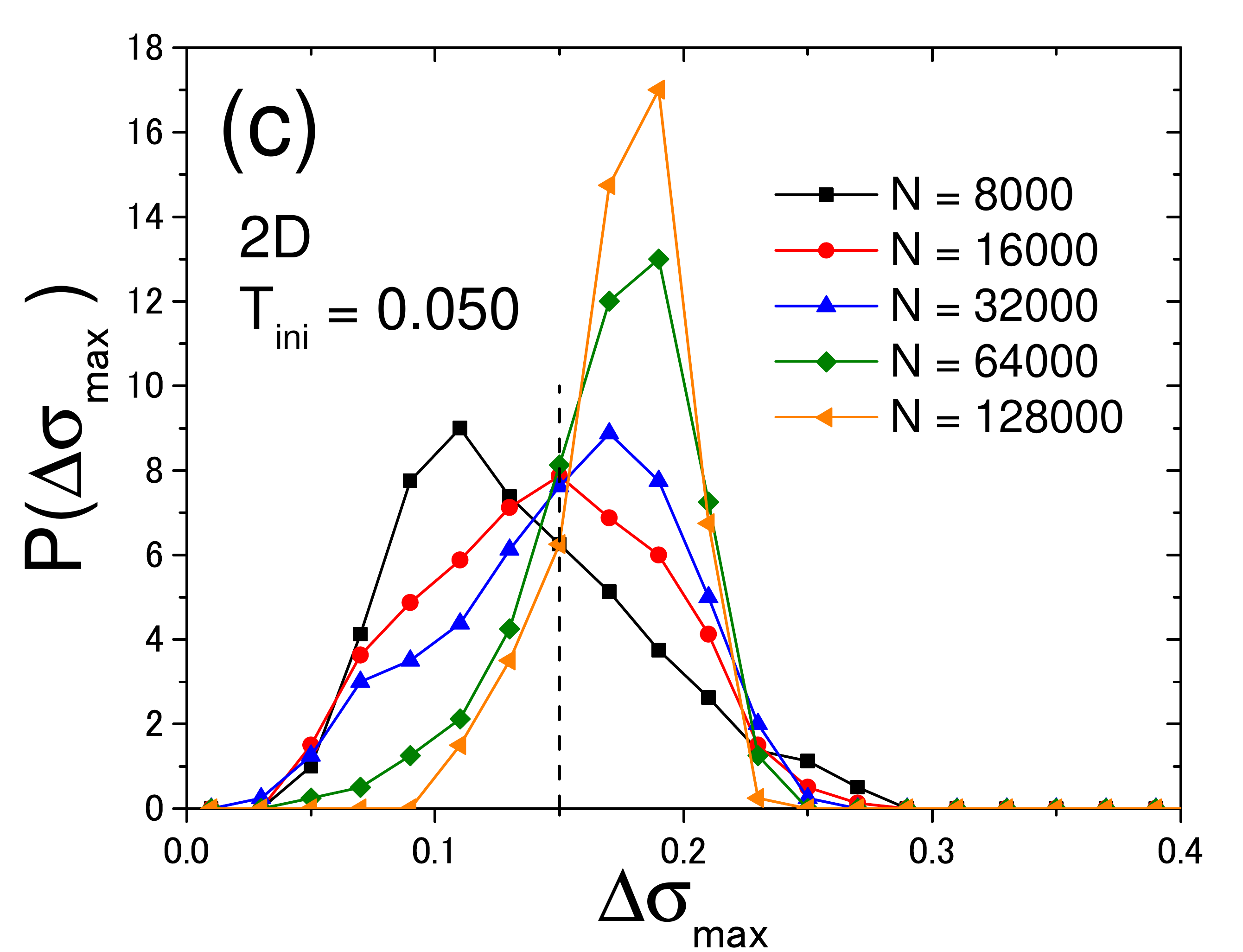}
\includegraphics[width=0.48\columnwidth]{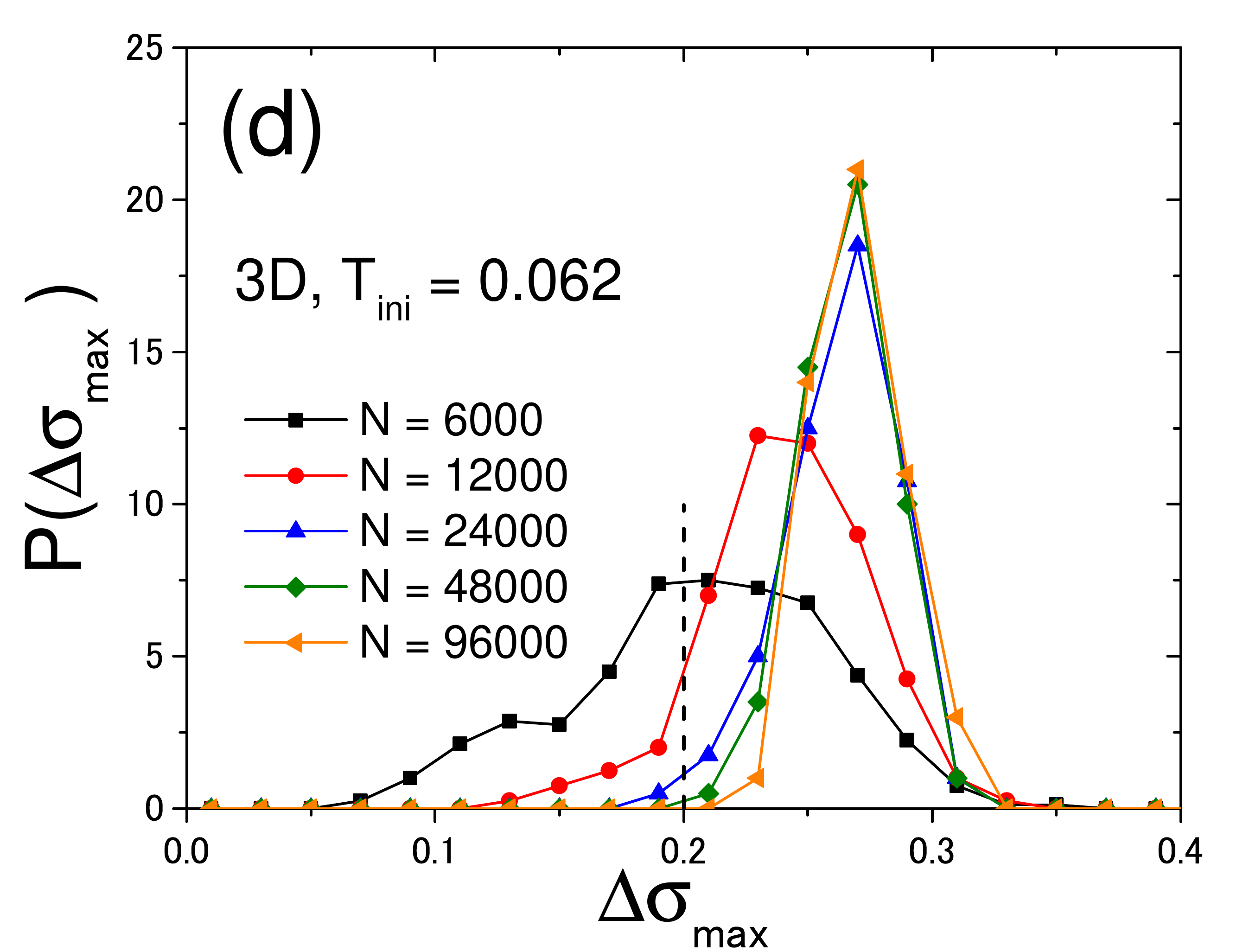}
\caption{Probability distribution of the maximum stress drop $\Delta \sigma_{\rm max}$ at $T_{\rm ini}=0.035$ (a), $0.040$ (b), and $0.050$ (c) in 2D. The 3D model at $T_{\rm ini}=0.062$ is shown in (d) for comparison. The vertical dashed lines indicates the chosen cutoff separating typical and anomalous samples.}
\label{fig:P_Delta_simga_max}
\end{figure}

\section{Identification of the anomalous samples} 
\label{sec:appendix_anomalous}

Figure~\ref{fig:P_Delta_simga_max} shows the probability distribution function of the maximum stress drop $\Delta \sigma_{\rm max}$ for 2D (a,b,c) and 3D (d) for low $T_{\rm ini}$'s. For the most stable case in 2D, $T_{\rm ini}=0.035$, there are two peaks for the smaller values of $N$. The right peak correspond to samples with a single large discontinuous stress drop, which we call typical samples,  and the left peak to samples with multiple stress drops at yielding, which we call anomalous samples. For  $N=8000$ the peak at smaller $\Delta \sigma_{\rm max}$ is dominant, which means that the majority of samples show multiple stress drops. However, the peak at higher $\Delta \sigma_{\rm max}$ grows with increasing $N$, and for large enough system size, most of the samples show a single large discontinuous stress drop at yielding (hence the denomination of typical samples); yet there remains a tail at smaller $\Delta \sigma_{\rm max}$ corresponding to the anomalous samples. We observe the same trend up to $T_{\rm ini}=0.050$, but the peak positions shift toward smaller $\Delta \sigma_{\rm max}$ with increasing $T_{\rm ini}$. Above this $T_{\rm ini}=0.050$, we do not find any hint of two separate peaks, which forbids any sensible distinction of typical and anomalous samples.

To separate anomalous samples from typical samples for $T_{\rm ini}=0.035-0.050$, we choose a cutoff $\Delta \sigma_{\rm max}=0.15$ which seems to reasonably distinguish anomalous samples ($\Delta \sigma_{\rm max} \leq 0.15$) from typical ones ($\Delta \sigma_{\rm max} > 0.15$): see Fig.~\ref{fig:P_Delta_simga_max} (a,b,c). There is some leeway in defining this cutoff value, but the conclusion in the main text does not change if we slightly change the value. In contrast to 2D, 3D systems do not show a clear bimodal distribution, as seen in Fig.~\ref{fig:P_Delta_simga_max} (d). Besides, the tail at smaller $\Delta \sigma_{\rm max}$ is significantly suppressed compared to 2D. To nonetheless make an attempt to quantify the fraction of anomalous samples, we have chosen a cutoff at $\Delta \sigma_{\rm max}=0.2$.

\begin{figure}
\includegraphics[width=0.48\columnwidth]{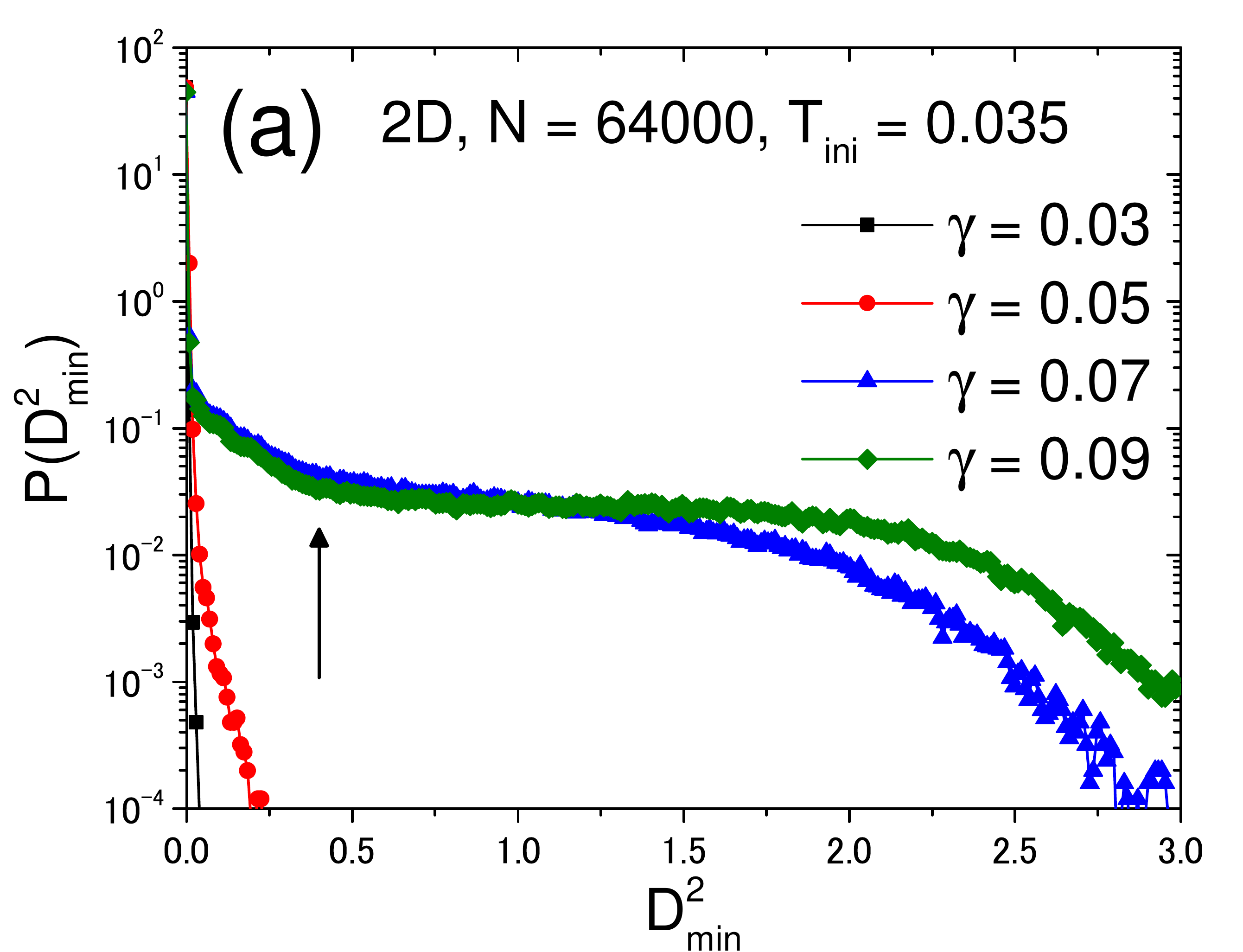}
\includegraphics[width=0.48\columnwidth]{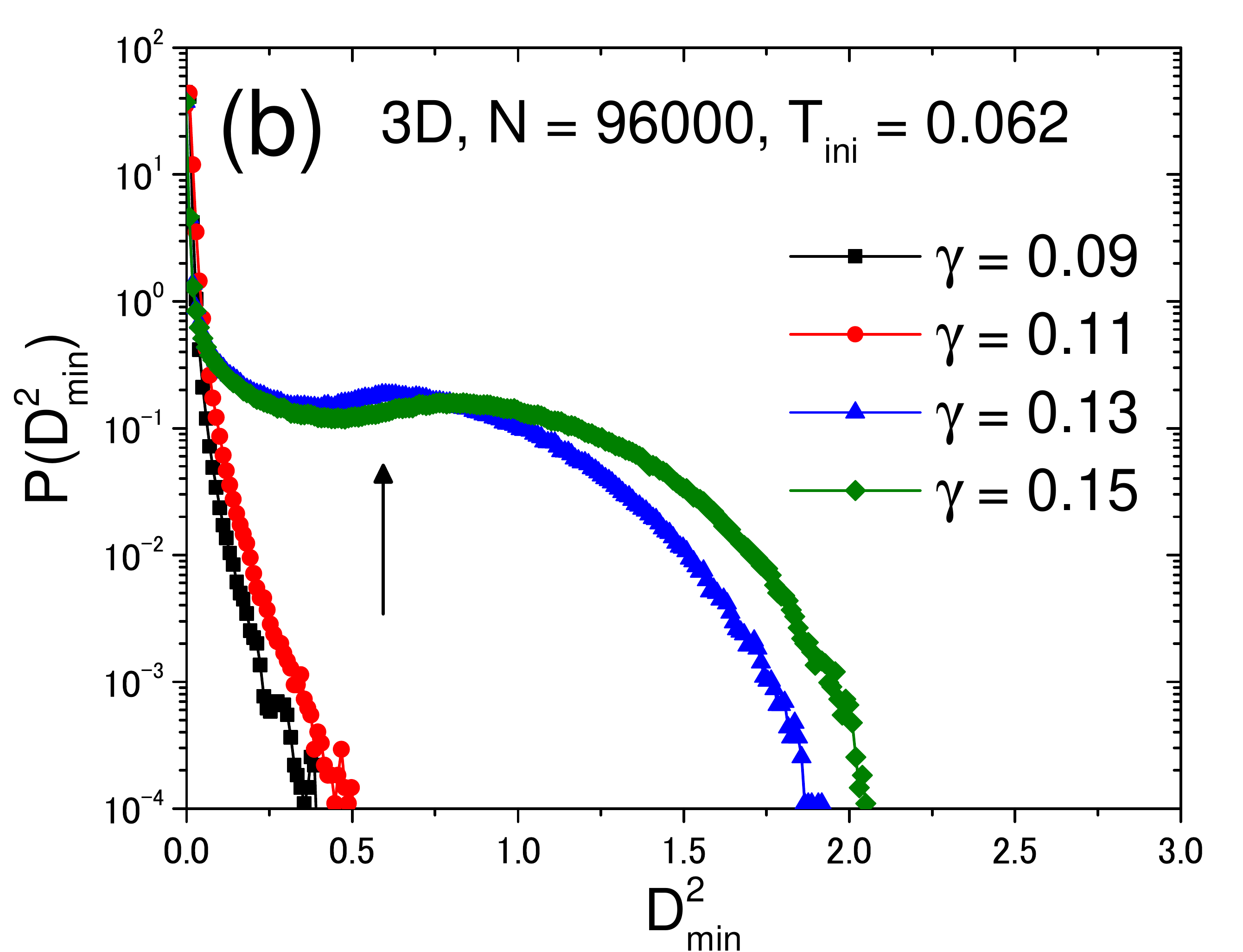}
\caption{
Probability distribution of the particle nonaffine (squared) displacement $D_{\rm min}^2$ in 2D (a) and in 3D (b) for several values of the strain $\gamma$ covering the regimes before and after yielding. The vertical arrows indicate the threshold above which particles are considered as belonging to a shear band: We have chosen $D_{\rm min}^2 >0.4$ in 2D and  $D_{\rm min}^2 >0.59$ in 3D.
}
\label{fig:P_Dmin2}
\end{figure}  

\begin{figure}[b]
\includegraphics[width=0.48\columnwidth]{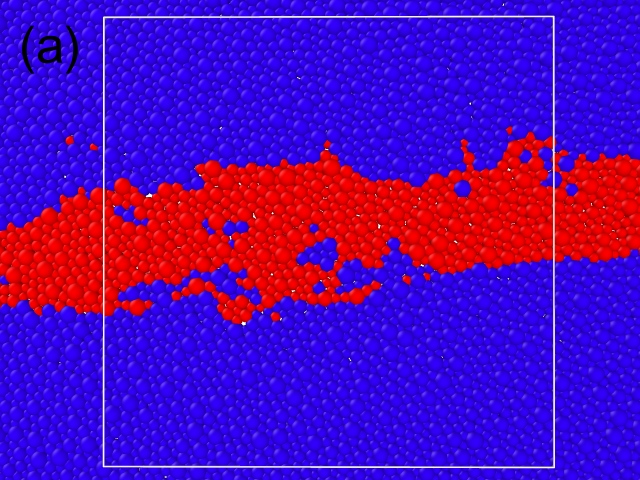}
\includegraphics[width=0.48\columnwidth]{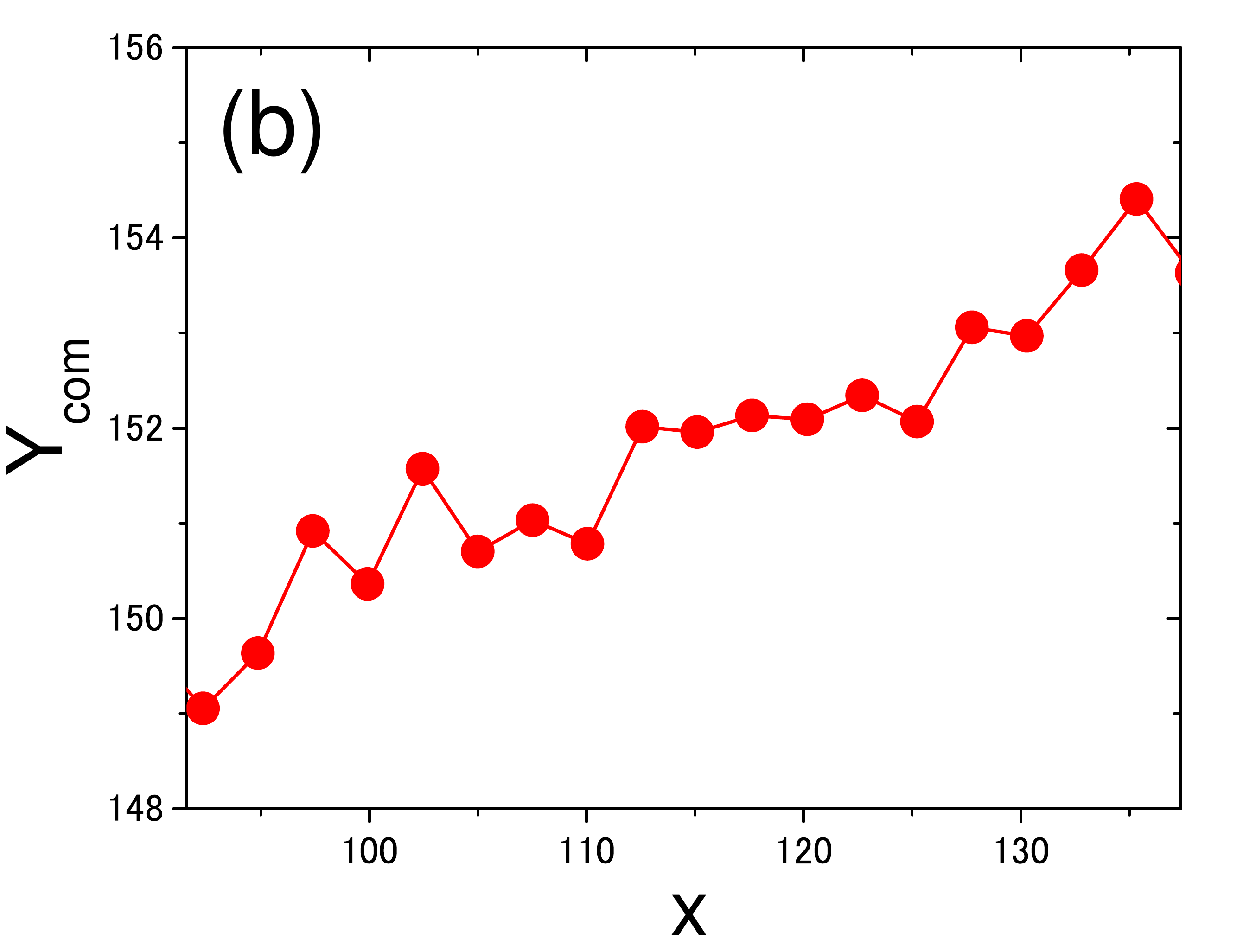}
\caption{
(a): Snapshot representing the particles belonging to a shear band (shown in red) for a 2D sample at a strain $\gamma=0.07$. The rest of the sample is essentially an elastic body and is shown in blue.
(b): The $y$-coordinate of the center of mass of the shear band for each bin along the $x$-direction.
}
\label{fig:trajectory_SB}
\end{figure}

\section{Shear band width} 
\label{sec:shear_band_width}

\begin{figure*}
\includegraphics[width=0.66\columnwidth]{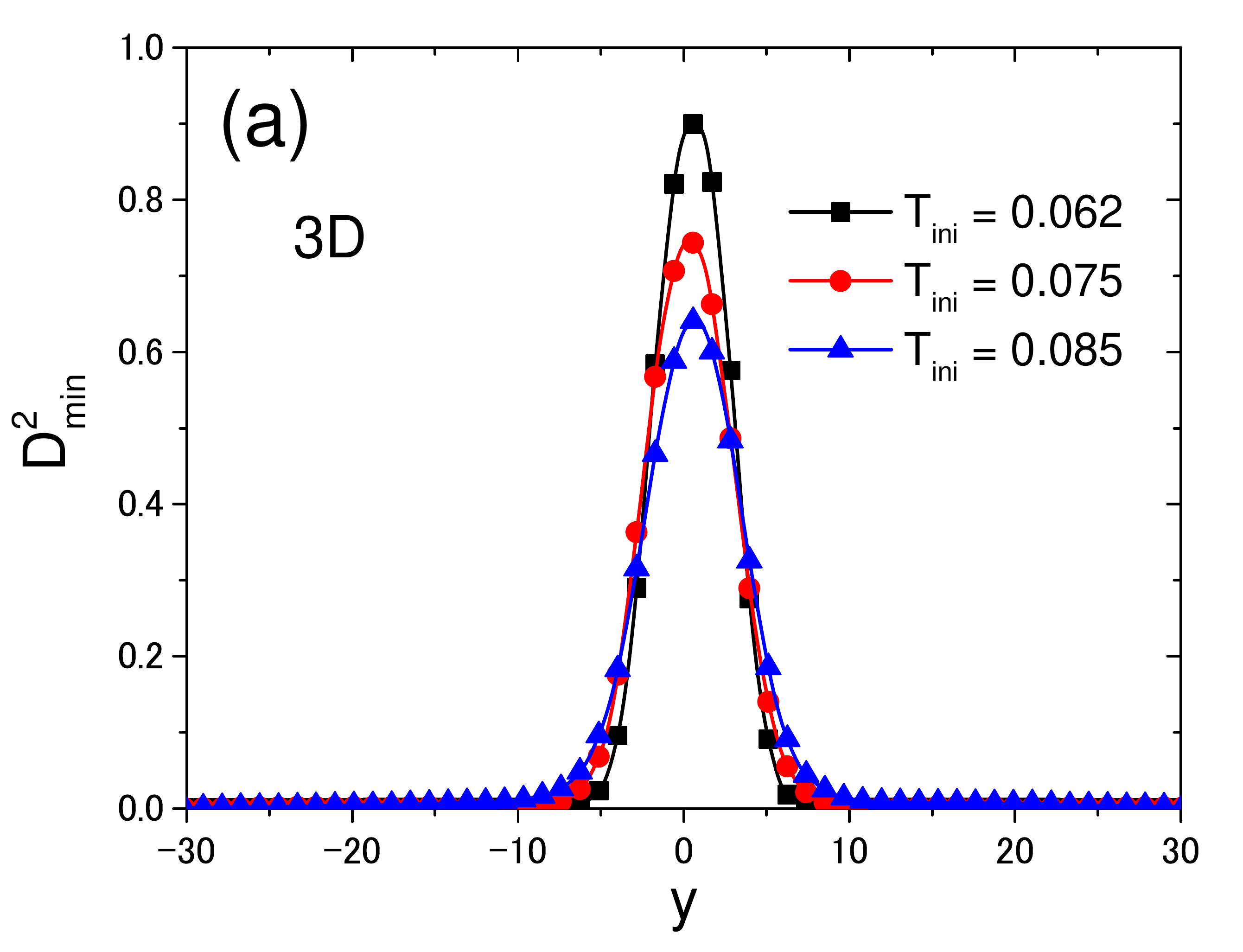}
\includegraphics[width=0.66\columnwidth]{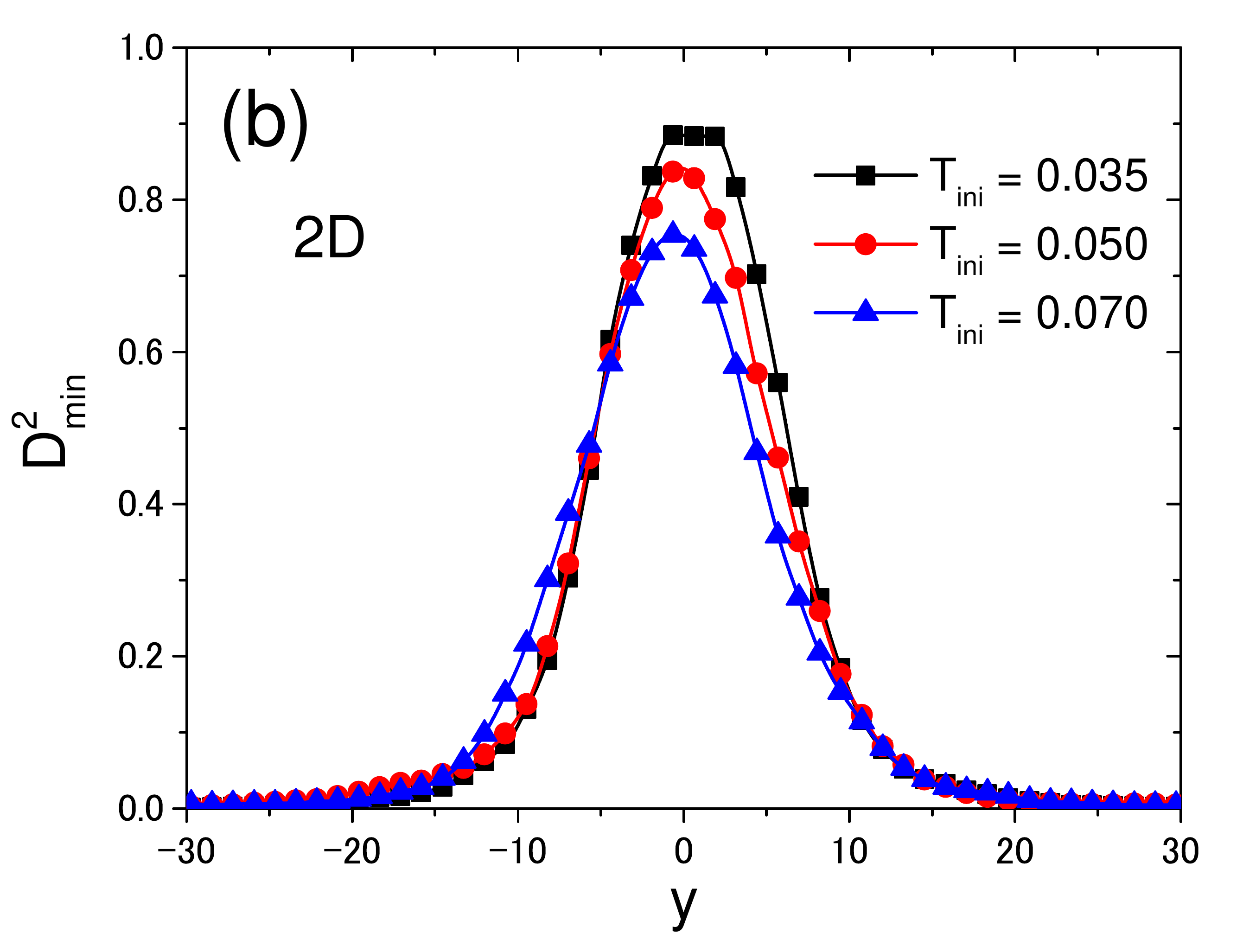}
\includegraphics[width=0.66\columnwidth]{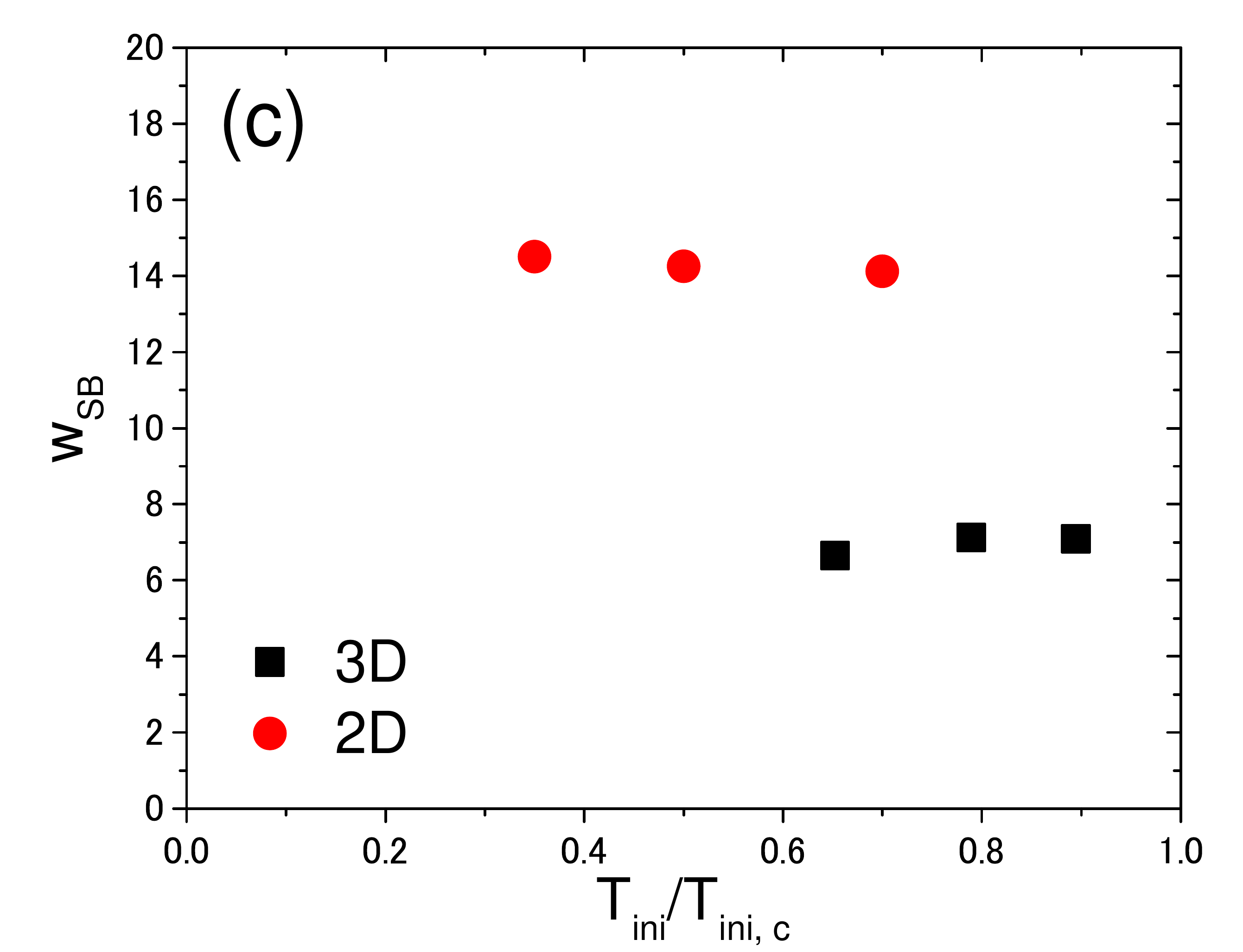}
\caption{
Averaged $D_{\rm min}^2$ profile along the direction perpendicular to the shear band for several preparation temperatures in 3D (a) and 2D (b).
(c): Width of the shear band as a function of the normalized preparation temperature $T_{\rm ini}/T_{\rm ini,c}$.
}
\label{fig:SB_width}
\end{figure*}

We measure a typical width of the shear band, $w_{\rm SB}$, and its temperature evolution, following a similar method conducted in Refs.~\cite{parmar2019strain,golkia2020flow}. We first divide the configuration into slabs along the direction perpendicular to the shear band, and then compute the average non-affine displacement $D_{\rm min}^2$ for each slab. $D_{\rm min}^2$ is computed between the origin $\gamma=0$ and $\gamma=0.07$ ($0.13$) for 2D (3D). Figures~\ref{fig:SB_width}(a, b) show the $D_{\rm min}^2$ profile obtained in 3D and 2D for several preparation temperature $T_{\rm ini}$. Clearly, the 2D systems have a wider $D_{\rm min}^2$ profile, in accord with the visual impression given by in Figs. 3(a,b) of the main text. Moreover, the width of the profile does not change so much along $T_{\rm ini}$ in both 3D and 2D. To quantify this feature, we operationally define $w_{\rm SB}$ as the width of the $D_{\rm min}^2$ profile at $D_{\rm min}^2=0.3$. We plot $w_{\rm SB}$ for several degrees of stability in Fig.~\ref{fig:SB_width}(c), where $T_{\rm ini}$ is normalized by the critical preparation temperature $T_{{\rm ini},c}$ to allow a comparison between 2D and 3D cases. We find that the width $w_{\rm SB}$ in 2D is always wider than that in 3D. This conclusion does not change when considering a different normalization temperature, e.g., an estimated experimental glass transition temperature, $T_{\rm g}$.

\section{Roughness analysis} 
\label{sec:appendix_rough}

We present some details on how we have computed the height-height correlation function $C(\Delta x)$ for the shear bands from the molecular simulation data. 

Particles are considered as part of the shear band if their nonaffine (squared) displacement $D_{\rm min}^2$ is large enough. Figure~\ref{fig:P_Dmin2} shows the probability distribution of $D_{\rm min}^2$, $P(D_{\rm min}^2)$, for several values of the strain $\gamma$ covering the regimes before and after yielding. Before yielding ($\gamma=0.03-0.05$ for 2D and $\gamma=0.09-0.11$ for 3D), $P(D_{\rm min}^2)$ is localized near the origin, which reflects the fact that most of the particles show a purely affine deformation and that only a very small fraction of particles undergo nonaffine displacements in the pre-yield regime. After yielding ($\gamma=0.07-0.09$ for 2D and $\gamma=0.13-0.15$ for 3D) on the other hand, a significant tail suddenly appears due to strain localization in the form of shear bands, and this tail grows with increasing $\gamma$. By introducing the thresholds shown in the vertical arrows in Fig.~\ref{fig:P_Dmin2}, we separate particles belonging to the shear band (with a nonaffine displacement above threshold, $D_{\rm min}^2>0.4$ for 2D and $D_{\rm min}^2>0.59$ for 3D) from particles undergoing affine displacement characteristic of a purely elastic solid. An illustration is given in Fig.~\ref{fig:trajectory_SB}(a) for a 2D sample.

We compute the average location of the shear band  (line in 2D or surface in 3D) by discretizing the base space ($x$ for a horizontal shear band in 2D, and ($x$,$z$) for a horizontal band in 3D). More specifically, we compute the $y$-coordinate of the center of mass, $Y_{\rm com}(x)$, of the particles belonging to the shear band and located within the bin specified by the position $x$ (or $x$ and $z$ in 3D). The output is illustrated in Fig.~\ref{fig:trajectory_SB}(b). For the bin width we have used $2.53$ in 2D and $2.28$ in 3D.

The height-height correlation functions $C(\Delta x)$ is finally defined as
\begin{equation}
C(\Delta x) = \langle (Y_{\rm com}(x+ \Delta x)-Y_{\rm com}(x))^2 \rangle_{x}^{1/2}
\end{equation}
in 2D. In 3D where the $z$-axis has to be taken into account, the averaging procedure for $C(\Delta x)$ is also performed along the $z$-direction, according to
\begin{equation}
C(\Delta x) =  \langle (Y_{\rm com}(x+ \Delta x, z)-Y_{\rm com}(x, z))^2 \rangle_{x, z}^{1/2}.
\end{equation}


%

\end{document}